\documentclass[10pt]{article}
\usepackage[margin=1.2in]{geometry}

\usepackage{Math598}
\usepackage{listings}
\usepackage{numprint}
\usepackage{enumerate}
\usepackage{bbm}
\usepackage{xcolor}
\usepackage{amsfonts,amsmath}
\usepackage{graphicx}
\usepackage{cancel}
\usepackage{url}
\usepackage{bbm}
\usepackage{cancel}
\usepackage{breqn}
\usepackage{booktabs}
\usepackage{array}
\usepackage{multirow}
\usepackage{fancyref}
\usepackage{hyperref}
\usepackage{rotating}
\usepackage{placeins}
\usepackage{verbatim}

\usepackage[numbers]{natbib}

\usepackage{amsthm}
\usepackage{animate}
\usepackage{eqnarray}

\usepackage[]{algorithm2e}
\usepackage{setspace}

\definecolor{keywordcolor}{rgb}{0,0.6,0.6}
\definecolor{delimcolor}{rgb}{0.461,0.039,0.102}
\definecolor{Rcommentcolor}{rgb}{0.101,0.043,0.432}

\lstdefinestyle{Rsettings}{
basicstyle=\ttfamily,
breaklines=true,
showstringspaces=false,
keywords={if, else, function, theFunction, tmp}, % Write as many keywords
otherkeywords={},
commentstyle=\itshape\color{Rcommentcolor},
keywordstyle=\color{keywordcolor},
moredelim=[s][\color{delimcolor}]{"}{"},
}

\def\T{\transpose}

\def\E{\mathbb{E}}

\def\one{\mathbbm{1}}

\def\calA{\mathcal{A}}

\def\za2{z_{\alpha/2}}

\def\xbf{\pmb{x}}

\def\alphabf{\pmb{\alpha}}
\def\betabf{\pmb{\beta}}
\def\psibf{\pmb{\psi}}

\def\Xbf{\pmb{X}}
\def\Ybf{\pmb{Y}}
\def\Abf{\pmb{A}}

\def\({\left(}
\def\){\right)}
\def\[{\left[}
\def\]{\right]}

\def\opt{\text{opt}}
\def\argmax{\text{argmax}\,\,\,}

\def\Resp{\text{Resp}}

\def\prev{\text{prev}}
\def\init{\text{init}}

\def\b{\\[1mm]}

\theoremstyle{definition}

\newtheorem{example}{Example}[section]

\begin{document}

\onehalfspacing

\title{Evaluating the Use of Generalized Dynamic Weighted Ordinary Least Squares for Individualized HIV Treatment Strategies}
\author{Larry Dong$^{1,2}$, Erica E. M. Moodie$^1$, Laura Villain$^2$\\and Rodolphe Thi{\'{e}}baut$^2$}
\date{%
    \small{$^1$Department of Epidemiology, Biostatistics and Occupational Health,
McGill University}\\%
    \small{$^2$University of Bordeaux, INSERM U1219 Bordeaux Population Health, Inria SISTM}
}
\maketitle

\begin{abstract}
Dynamic treatment regimes (DTR) are a statistical paradigm in precision medicine which aim to optimize patient outcomes by individualizing treatments. At its simplest, a DTR may require only a single decision to be made; this special case is called an individualized treatment rule (ITR) and is often used to maximize short-term rewards. Generalized dynamic weighted ordinary least squares (G-dWOLS), a DTR estimation method that offers theoretical advantages such as double robustness of parameter estimators in the decision rules, has been recently extended to now accommodate categorical treatments. In this work, G-dWOLS is applied to \textit{longitudinal} data to estimate an optimal ITR, which is demonstrated in simulations. This novel method is then applied to a population affected by HIV whereby an ITR for the administration of Interleukin 7 (IL-7) is devised to maximize the duration where the CD4 load is above a healthy threshold (500 cells/$\mu$L) while preventing the administration of unnecessary injections.
\end{abstract}

\section{Introduction} In precision or personalized medicine, the central paradigm lies in adopting patient-centric medical practices rather than disease-centric approaches \cite{ramaswami2018precision}. Because people are intrinsically different, the effects of available treatments can be highly variable from one patient to another. Particularly in the clinical management of chronic illnesses, the evolving nature of ailments such as cancer, depression and substance abuse amongst many other long-term conditions often calls for treatments to be adapted to patient response, characteristics and overall well-being \cite{chakraborty2013statistical, ramaswami2018precision}.\b

Dynamic treatment regimes (DTR) are a statistical paradigm in precision medicine that optimizes an outcome of interest through sequential decision making. In general, the procedure of optimizing for multi-stage decision problem requires knowledge of the underlying data-generating mechanism. Although the mechanism is often unknown, it -- or components of it -- can be estimated using data-driven methods. This is where evidence-based medicine meets statistics, where the development of a reliable theoretical framework is fundamental in finding optimal treatment regimes and in quantifying their uncertainty. The primary goals of DTR are twofold: comparing expected utilities of different deterministic treatments and obtaining optimal individualized treatment plans for patients \cite{chakraborty2013statistical}. Formally speaking, a DTR is a function that receives patient history as input and outputs an optimal decision vector which itself is comprised of one or multiple decisions, depending on the nature of the problem. At its simplest, a DTR is an estimation procedure that optimizes for a single decision; such single-stage DTR are referred as individualized treatment rules (ITR). One of the biggest strengths of the DTR framework is its ability to optimize long-term outcomes which involve medical interventions that are performed in a cascading fashion.\b% In a multi-stage setting, the set of decision rules can be perceived as a medical protocol which can be adapted by monitoring patient well-being as they progress through a specific treatment strategy. 

The primary objective of this current work is to demonstrate how generalized dynamic weighted ordinary least squares (generalized dWOLS or G-dWOLS), a DTR estimation method, can be used to optimize short-term or immediate outcomes from longitudinal data; we refer to such treatment regimes as myopic due to the nearsightedness of the maximization procedure. Although DTR is known for its ability to optimize multi-stage decision problems, we illustrate that developing myopic treatment regimes can be done for problems where it is reasonable to assume no delayed treatment effects. While such global optimal treatment strategies are preferable to locally optimal regimes, the benefits in solving for myopic treatment regimes are its simpler interpretation and a more lightweight statistical estimation procedure. By performing this methodological simplification, we remove the recursive process in the estimation of an optimal multi-stage DTR \cite{chakraborty2013statistical}. The underlying correlation between observations from a same individual can be accounted for by applying G-dWOLS within the generalized estimating equation (GEE) framework. We can accommodate the correlation between observations in the same individuals in a linear modelling framework; this works well with the G-dWOLS approach since it solves for parameters of interest in a weighted least squares (more details in section~\ref{subsection:dwols}).\b

This methodological extension of G-dWOLS is motivated by the therapeutic benefits of Interleukin 7 (IL-7) injections in low immunological responders - HIV-infected individuals who are characterized by a low CD4 count despite a lack of viral load following antiretroviral treatment \cite{thiebaut2014quantifying}. Clinical trials such as INSPIRE 2 and 3, whose protocol is available in Appendix A of the Supplementary Material, have been conducted to evaluate the effect of exogenous IL-7 injections on the immune system in low immunological responders who have been treated with antiretrovirals \cite{levy2012effects, thiebaut2014quantifying}. Using a simulation study available in Appendix B of the Supplementary Material, we show that the theoretical properties of G-dWOLS still hold when devising an ITR with longitudinal data and multiple treatment categories. An application of our proposed method is then performed on data from INSPIRE clinical trials to see if tailoring treatment with respect to subject-level information would be clinically relevant.\b

Following the presentation of the INSPIRE data in section 2, a summary of our proposed method are stated in section 3. In section 4, we explain how G-dWOLS will be applied to data from the INSPIRE studies and data analysis results are elaborated in section 5. Lastly, the benefits and limitations of the work are considered in section 6.

\section{INSPIRE Trials and Related Work} HIV is characterized by a depletion of CD4+ T cells (CD4), which are a pillar of the immune system (\cite{douek2009emerging}). Highly active antiretroviral therapy (HAART), the reference treatment for HIV, inhibits the replication of the virus and it is most often followed by an increase of CD4 T cells (\cite{douek2009emerging}). It has been shown that failure to reconstitute of CD4 cells in HIV-infected individuals to a threshold greater or equal to 500 cells/$\mu$L is associated with a higher mortality rate and an increased risk in developing opportunistic infections (\cite{lewden2007hiv, logerot2018hiv, opportunistic2012cd4}). However, despite having no detectable viral load, 15\% to 30\% of individuals receiving HAART are known as poor or low immunological responders due to their inability to increase their CD4 counts and hence reconstitute their immunological system (\cite{logerot2018hiv, grabar2000clinical}).\b

Clinical trials INSPIRE 1, 2 and 3 were conducted to investigate the benefits of administering injections of IL-7 in low immunological responders by attempting to increase participants' CD4 count above 500 cells/$\mu$L \cite{levy2009enhanced, thiebaut2016repeated}. IL-7 is a cytokine that plays an essential role in the survival and maintenance of CD4 cells \cite{lawson2015interleukin, surh2005regulation, surh2008homeostasis, thiebaut2014quantifying}. While this protein is naturally produced in stromal and epithelial cells of the bone marrow and thymus, multiple studies conducted both in humans and in animals suggest that the administration of IL-7 leads to the reconstitution of the immune system \cite{mackall2011harnessing, rosenberg20067, sportes2008administration, surh2008homeostasis}. The main finding of INSPIRE 1 was that, after receiving injections of IL-7, participants exhibited improvements of many immunological markers including a sustained increase of CD4 cells \cite{thiebaut2016repeated}. However, a gradual decrease in CD4 counts followed which motivated the assessment of repeated cycles of IL-7 injections in INSPIRE 2 and 3. Three doses of IL-7 (10, 20 and 30 $\mu$g/Kg) were used in INSPIRE 1 whereas injections in INSPIRE 2 and 3 were all of 20 $\mu$g/Kg dosage due to its compromise in treatment effectiveness and side effects \cite{levy2009enhanced, villain2019adaptive}. In this study, only data from INSPIRE 2 and 3 were used due to the lack of baseline patient characteristics in INSPIRE 1.\b

INSPIRE 2 was a single-arm clinical trial where participants were drawn from an adult population of people living with HIV who have been receiving HAART for over a year and exhibiting suboptimal CD4 counts, i.e.~between 100 and 400 cells/$\mu$L \cite{jarne2017modeling, levy2012effects}. The protocol entails providing three injections of IL-7 at one week intervals, monitoring patient T-cell response quarterly and providing another set of injections after 12 months of follow-up only if their CD4 count fell below 550 cells/$\mu$L \cite{levy2012effects, thiebaut2016repeated}. INSPIRE 3 was a clinical study where participants were randomized into a IL-7 arm and a control arm at a 3:1 ratio. With eligibility criteria defined similarly to those used in INSPIRE 2, the clinical protocol in the IL-7 arm entailed beginning patients with three injections, and readministrating three injections if patients presented a CD4 count $<$ 550 cells/$\mu$L at any quarterly evaluations. Patients in the control arm had their CD4 count measured for one year without any IL-7 injections. After one year, similarly to participants in the IL-7 arm, injections were administered if patients in the control arm presented a CD4 count below 550 cells$/\mu$L \cite{thiebaut2016repeated}. The discrepancy between the protocol CD4 threshold for treatment administration of 550 cells/$\mu$L and ``healthy'' threshold of 500 cells/$\mu$L accounts for possible measurement error and an anticipated immediate decrease in CD4 counts below 500 cells/$\mu$L if participants exhibited CD4 levels between 500 and 550 cells/$\mu$L \cite{thiebaut2014quantifying, thiebaut2016repeated}. From INSPIRE 2 and 3, a combined total of 113 patients received 198 sets of injections; two participants withdrew from the study. Over the course of the study, 1300 adverse events related to the administration of IL-7 have been recorded, although most of them were not deemed serious \cite{thiebaut2016repeated}. Of all reported occurrences, 77.6\% were grade $\leq$ 1, 20.7\% were grade 2 and 1.7\% were grade $\geq$ 3; three patients experienced serious side effects and there was no significant difference in the number of adverse events for each visit \cite{thiebaut2016repeated}. A figure visualizing the INSPIRE protocols is available in Appendix A of the Supplementary Material.\b

The goal of the data analysis presented in section~\ref{section:data-analysis} is to devise a myopic treatment rule that optimizes the number of injections while preventing the administration of unnecessary ones using G-dWOLS \cite{petersen2007individualized}. There are costs -- both financial and clinical (such as treatment fatigue and risk of adverse effects) -- to injections, and so it is of interest to find the smallest number of injections needed to ensure that CD4 concentration lies above 500 cells/$\mu$L.% Such adaptive strategies tailor treatment to patient characteristics to maximize short-term patient outcomes as the benefits of IL-7 injections on the increase in CD4 count appear to be immediate. 
 
\section{ITR Estimation for Longitudinal Data} % 2

ITR can be used in the optimization of long-term outcomes through the maximization of short-term or immediate rewards. The overarching idea behind this approach bears strong similarities with the greedy algorithm whereby locally optimal decisions are made in the attempt to maximize an overall or terminal outcome \cite{kleinberg2006algorithm}. The essence of this method stems from the trade-off between the simpler formulation of the problem-solving framework and a potentially non-optimal solution that is relatively close to the best one \cite{kleinberg2006algorithm}.\b

The process of estimating optimal ITR, also known as myopic optimal rules, can involve less model-based extrapolation while yielding more precise (i.e.~narrower) confidence intervals \cite{petersen2007individualized}. Rather than evaluating regimes through a sequence of decision rules, one of many ways to approach the study design to treat every stage as a single observation. In doing so, the sequential nature of interventions is eliminated and the optimization procedure ensues with the maximization of short-term rewards.\b

Of course, there are many examples where a greedy algorithm or a myopic treatment strategy may fail to provide a reasonable policy to optimize long-term outcomes. In  Appendix C in the Supplementary Material), we outline a sufficient set of conditions -- requiring assumptions on the immediacy of the effect of treatments and potential interactions between treatments over time -- that are needed to suggest the myopic strategy is adequate, and relate these to the INSPIRE setting, which is the focus of this work.\b

%This section on statistical methodology will be divided into two parts. In a first instance, we provide an overview of the DTR framework by focusing on single-stage decision problems, i.e. ITR, and summarize the G-dWOLS estimation procedure. In a second instance, we discuss the generalizing estimation equation (GEE) framework and how we incorporate such methods for handling longitudinal data in ITR estimation using G-dWOLS. 

\subsection{Overview of Regression-Based Approaches for ITR Estimation}
\label{subsection:overview}

When working with ITR, positing simpler models such as linear regressions to estimate contrasts of utilities offers simpler interpretability of estimated parameters. A \textbf{blip} function denoted by $\gamma(\xbf, a; \psibf)$ is defined to be the expected gain in outcome if treatment $A = a$ were to be chosen instead of a reference or baseline treatment $a^{\text{ref}}$. For instance, in the binary case where $A \in \{\text{0, 1}\}$, $a^{\text{ref}} =$ 0 often represents an absence of treatment. Non-parametrically, this is given by
  \begin{align*}
    \gamma(\xbf, a; \psibf) = \E\[Y\(\xbf, a\) - Y\(\xbf, a^{\text{ref}}\) \, | \, \Xbf = \xbf, A = a\]
  \end{align*}
where $Y(\xbf, a)$ denotes the counterfactual outcome under treatment $a$ with covariates $\xbf$. In other words, the blip function characterizes the expected gain in the outcome upon providing treatment $A = a$ compared to the reference treatment $a^{\text{ref}}$. When modelling for the outcome variable $Y$, the model is often written as the sum of two expressions: the treatment-free model and the blip function, which are parametrized by the respective coefficients $\betabf$ and $\psibf$. The \textbf{treatment-free} model $G(\xbf; \betabf)$ is the portion of the outcome model that is independent of the treatment:

\begin{align*}
    \underbrace{\E\[Y\,|\,\Xbf=\xbf, A=a; \betabf, \psibf\]}_{\text{outcome model}} = \underbrace{G(\xbf; \betabf)}_{\text{treatment-free model}} + \underbrace{\gamma(\xbf, a; \psibf)}_{\text{blip function}}.
\end{align*}

%At its simplest, we have that $\E\[Y\,|\,\Xbf=\xbf, A=a; \betabf, \psibf\] = \xbf^\beta \betabf + a\cdot \xbf^\psi \psibf$ where $A \in \{\text{0, 1}\}$ is a binary treatment variable and $\xbf^\beta, \xbf^\psi$ are (potentially identical) subset of covariates $\xbf$. Another way to understand the treatment-free model is that it is the expected outcome under baseline treatment, since, $\gamma(\xbf^\psi, a^{\text{ref}}; \psibf) =$ 0 by definition. 
The nomenclature for the models in the outcome model is due to the clear separation of linear expressions due to the terms which interact with the treatment variable of interest. In the binary case where $A \in \{0, 1\}$, the blip function can be understood as an expected gain in utility or reward when receiving treatment $A =$ 1 compared to $A =$ 0; it is assumed that $A =$ 0 is the baseline or reference treatment. However, when more than 2 treatments are possible, adaptations to blip function and hence the estimating function must be made. Keeping the treatment-free model untouched, the sum of $m-1$ contrast terms will comprise the ``new'', multi-treatment blip function. In other words, a linear term $\xbf^\psi \psi_{\ell}$ must be posited for each non-reference treatment $a_\ell$. When $A$ is a categorical variable where $\mathcal{A} \equiv \{a_0, a_1, \dots, a_{m-1}\}$, assume without loss of generality that $a_0$ is chosen to be the reference treatment. The blip function can be written as a sum of treatment contrasts, each of which represents an expected gain in utility compared to the counterfactual situation in which $a_0$ was the assigned treatment:

\begin{align*}
  \gamma(\xbf, a; \psibf) = \sum_{\ell=1}^{m-1} \one_{a = a_\ell}\, \xbf^{\psi}\psibf_\ell
\end{align*}
where $\one_{a = a_\ell} = 1$ if $a = a_\ell$. Linear contrast functions $\gamma_\ell(\cdot)$ are often posited for convenience, but, in theory, any functional form which equals 0 when evaluated at the reference treatment is permissible. ITR estimation methods such as G-dWOLS also calls for a treatment model, also referred as (generalized) propensity score or conditional treatment density function, denoted by $\pi(\xbf, a)$, which can be estimated using (multinomial) logistic regression \cite{austin2018assessing, rosenbaum1983central}:

\begin{align*}
    \pi(\xbf, a; \alphabf) = \E\[A = a \, | \, \Xbf = \xbf; \alphabf\] \, .
\end{align*}

The treatment model parameters are estimated separately, then weights are constructed and plugged into the regression used to estimate outcome model parameters in a two-step estimation procedure. Whether in a dynamic treatment regime setting or, as is our focus here, a repeated ITR setting where a myopic regime is estimated via longitudinal data, it is typical to model the propensity score in each treatment interval independently, without taking into account within-person correlation. Under the assumption of no unmeasured confounding, this approach is justified. In this work, the three models detailed above -- treatment, treatment-free and blip model -- are respectively parameterized by coefficients $\alphabf$, $\betabf$ and $\psibf$. Regression-based approaches for ITR estimation posit linear models for the latter two, i.e.~$G(\xbf; \betabf) = \xbf^\beta \betabf$ and $\gamma(\xbf, a; \psibf) = \xbf^\psi \psibf$. Superscripts $\beta$ and $\psi$ are used to label explanatory variables with respect to their respective ``submodel'' within the outcome model, whereby $\Xbf^\beta$ and $\Xbf^\psi$ are typically subsets of subject-level covariates denoted by $\Xbf$.% (and indeed, $\Xbf^\psi$ is often a subset of $\Xbf^\beta$). 

\subsection{Generalized Dynamic Weighted Ordinary Least Squares} % 2.3
\label{subsection:dwols}

When working with ITR, positing simple models to model contrasts of utilities offers straightforward interpretability of estimated parameters \cite{chakraborty2013statistical}. Recent work has examined the use of statistical learning models such as decision trees \cite{tao2018tree} and deep neural networks (\cite{liu2019learning}). However, we will not pursue such approaches but rather focus on G-dWOLS, a regression-based approach in the estimation of treatment regimes. Other related DTR estimation methods include Q-Learning \cite{murphy2005generalization, moodie2012q}, G-estimation \cite{robins2004optimal} and dynamic weighted ordinary least squares (dWOLS) \cite{wallace2015doubly}, the estimation method on which G-dWOLS builds upon Schultz and Moodie (2020) \cite{schulz2020doubly}.\b

The G-dWOLS method is a weighted regression-based approach for the estimation of optimal treatment regimes first introduced by \cite{wallace2015doubly} and further extended by Schultz and Moodie (2020)\cite{schulz2020doubly}. This subsection highlights the theoretical underpinnings and practical considerations detailed in the latter article. Recent work in the dWOLS literature is now also able to handle survival outcomes \cite{simoneau2019estimating} and continuous treatments \cite{schulz2020doubly}, but this work focuses on categorical treatments. Other DTR estimation procedures for categorical treatments include Qi et al. (2020) \cite{qi2020multi} and Xue et al. (2020) \cite{xue2020multicategory}. The main advantages of using G-dWOLS compared to other methods are threefold: its relatively intuitive implementation, its ability to accommodate categorical treatments and its statistically robust estimation procedure (more details in the next subsection). G-dWOLS provides doubly-robust blip parameters by relying on having weights $w(\xbf, a)$ satisfying the balancing property below. This estimation procedure also has the potential of being less daunting in implementation due to the familiar function form of estimating function, as it is equivalent to minimizing a weighted least squares. This equivalence in methodological procedure allows a relatively straightforward implementation of the G-dWOLS algorithm using built-in regression functions with weights $w(\cdot)$ such that $\pi\(\xbf, a\)w\(a, \xbf\) = \pi(\xbf, a')w\(a', \xbf\)$ is satisfied for all $a, a' \in \calA$ where $\pi\(\xbf, a\) = P\(A = a\,|\,\Xbf = \xbf\)$ \cite{schulz2020doubly}. This equality stems from the theorem of the balancing property in dWOLS and G-dWOLS literature and a detailed proof of this is provided in Wallace and Moodie (2015) \cite{wallace2015doubly} for binary treatments and Schultz and Moodie (2020)\cite{schulz2020doubly} for categorical treatments.\b

According to the balancing property, doubly-robust blip parameters are ensured by using correctly specified propensity scores in the weights associated to each observation in the weighted least squares minimizing. The formulation of the theorem is general, in that the product $\pi\(\xbf, a\)w\(\xbf, a\)$ needs to bear the same value for all treatments $a \in \calA$. Examples of multinomial weights adhering to the balancing property that will be discussed in this work include inverse probability of treatment (IPT), $w(\xbf, a) = \{\pi(\xbf, a)\}^{-1}$, and overlap weights, $w(\xbf, a) = \{\pi(\xbf, a)\cdot \sum_{\ell=1}^m {\pi(\xbf, a_\ell)}^{-1}\}^{-1}$. Notice that the overlap weights are in fact IPT weights divided by a stabilizing term $\sum_{\ell=1}^m {\pi(\xbf, a_\ell)}^{-1}$ that solely depends on $\xbf$. This dissimilarity in weight forms allows overlap weights to be bounded between 0 and 1, whereas IPT weights which are unbounded from above. In Schultz and Moodie (2020) \cite{schulz2020doubly}, other weight forms for categorical treatments adhering to the balancing property are also provided, but will not be addressed in this paper as they offer no discernible advantage over the IPT and overlap weights.\b

Without the balancing weights, the G-dWOLS estimation process simplifies to Q-learning in a single-stage setting \cite{wallace2015doubly}. In practice, when using G-dWOLS for ITR estimation, the selection of the weights comes down to potential gain in efficiency and researcher's preference. Standard error estimates can vary depending on the form of weights since the asymptotic variance of blip parameters depend on $w(\cdot)$ \cite{schulz2020doubly, wallace2015doubly}. The G-dWOLS estimation procedure for ITRs is described in the five following steps:

\begin{itemize}
  \item[1.] Select (possibly identical) subsets of covariates $\Xbf^\alpha, \Xbf^\beta$ and $\Xbf^\psi$ from $\Xbf$ for the treatment model, the treatment-free model and the blip function respectively. We let covariate matrices $\Xbf_i^\beta$ and $\Xbf_i^\psi$ be of $(p \times 1)$ and $(r \times 1)$ dimension.
  \item[2.] Propose a treatment model $\E\[A\,|\,\Xbf\]$ and define a weight $w$ such that the balancing condition is satisfied.
  \item[3.] Posit treatment-free model $f(\cdot)$ in $\E\[Y \,|\,\Xbf, A\] = f\(\Xbf^\beta \betabf\) + \sum_{\ell=1}^{m-1} \one_{A = a_\ell}\Xbf^\psi \psibf_{\ell}$. Typically, linear regressions are used due to their appealing statistical properties and simplicity, i.e.~$f(\Xbf^\beta, \betabf) = \Xbf^\beta \betabf$.
  \item[4.] Solve the following system of estimating equations to obtain the blip function parameter estimates $\widehat{\psibf}$:
  \begin{align*}
    \pmb{0}_{\big(p + (m - 1)r\big)\times 1} &= \sum_{i=1}^n
    \begin{pmatrix}
        \Xbf_i^\beta\\
        \one_{A_i=a_1}\Xbf_i^{\psi}\\
        \vdots\\
        \one_{A_i=a_{m-1}}\Xbf_i^{\psi}
    \end{pmatrix}
    w_i \(Y_i - \Xbf^\beta \betabf -  \sum_{\ell=1}^{m-1} \one_{A_i = a_\ell}\, \
    \Xbf^{\psi}\psibf_\ell \) \,.
    \end{align*}
    The solution to the score function is a $\big(p + (m-1)r\big) \times 1$-dimensional parameter vector: $\left(\betabf, \psibf_1, \dots, \psibf_{m-1})\right)$.
  \item[5.] Evaluate the optimal treatment plan for each subject given estimates $\widehat{\psibf}$ as follow:
  \begin{align*}
    \widehat{A}^\opt(\xbf^\psi) = 
    \begin{cases}
        \underset{\ell \in \{1, \dots, m-1\}}{\argmax}  \xbf^\psi\widehat{\psibf}_{a_\ell} & \qquad \text{if } \underset{\ell \in \{1, \dots, m-1\}}{\text{max}} \xbf^\psi\widehat{\psibf}_{a_\ell} > 0\\
        \,a_0 & \qquad \text{otherwise.}
    \end{cases}
\end{align*}
\end{itemize}

\subsection{Generalized Estimating Equation Framework}

When solving multi-stage decision problems using a myopic regime study design, each participant $i$ contributes $n_i$ observations into an agglomerated dataset; in a clinical context, each observation can be referred to as patient-stage. Because there are many observations for a single individual, an intuitive assumption is to suppose that there is some degree of correlation between measurements contributed by the same person. Many statistical concepts surrounding generalized estimating equations (GEE) have been developed to accurately estimate parameters while acknowledging underlying correlation structures. This general framework provided in \cite{diggle2002analysis} allows statistical inference to be performed on linear coefficients $\betabf$ and $\psibf = \left(\psibf_1, \dots, \psibf_{m-1}\right)$ under the assumption of different correlation structures for repeated measurements \cite{diggle2002analysis}. Notation and terminology are largely inspired from Chapter 4 in \cite{diggle2002analysis}.\b

To account for the correlation of repeated measurements, different structures for the variance matrix can be assumed $V_i = \text{Var}(\pmb{Y}_i)$, which is also referred as the working covariance matrix and may differ from the true variance structure \cite{diggle2002analysis}. The primary benefit in using a GEE framework is that the estimators of linear coefficients are asymptotically normal even when misspecifying the working covariance structure $V_i$ \cite{diggle2002analysis}. This is particularly useful in situations where the underlying correlation structure is unknown. Based on previous research \cite{wang2015perils, pels2018call}, we advocate for the use of an independence working correlation, however an autoregressive structure could also be a reasonable choice.\b

Although longitudinal data call for a GEE or similar  approach, in G-dWOLS, the procedure for estimating the standard errors of blip coefficients for an ITR also needs to incorporate the variability due to estimating generalized propensity scores (or parameters of a treatment model). The G-dWOLS method incorporates estimated rather than known weights into the estimation procedure of DTR. In practice, these weights are estimated values, which means that an adjustment to the robust standard error formula is required for them to provide valid inference. One way to circumvent this issue is to estimate standard errors via bootstrapping \cite{efron1992bootstrap}. In G-dWOLS, bootstrap estimation of standard errors can be beneficial in that it accounts for the uncertainty attributable to the estimation of the generalized propensity score. However, it is also prone to suffer from small sample size issues since resampling must be performed on individuals rather than observations to retain the underlying correlation structure due to repeated measurements across bootstrap iterations.

\subsection{Summary of Proposed Method}

Combining elements from G-dWOLS and methods to account for repeated measurements, an overview of the proposed method for ITR estimation thesis is provided below.\b

For each individual $i$ with $n_i \geq 1$ observations, given a sequence of outcome variables $\Ybf_i = \{Y_{i1}, \dots, Y_{in_i}\}^\T$, categorical treatment $\Abf_i = \{A_{i1}, \dots, A_{in_i}\}^\T$ with $A_{i\cdot} \in \calA \equiv \{a_0, \dots, a_{m-1}\}$ and time-varying subject-level characteristics $\Xbf_i = (\Xbf_{i1}, \dots, \Xbf_{in_i})^\T$, we wish to understand the treatment effects of each treatment option $a_j \in \calA$ on the outcome $Y$. We first concatenate all observations into a single set denoted by $\mathcal{D} = \left\{\left(Y_{ij}, A_{ij}, \Xbf_{ij} \right) \right\}$. As detailed in section~\ref{subsection:overview}, we propose the following linear model for the outcome variable:
\begin{align*}
    \E\left[Y_i\,|\,\Xbf_i, A_i; \betabf, \psibf\right] = \Xbf_i^\beta \betabf + \sum_{\ell=1}^{m-1} \one_{A_i = a_\ell}\Xbf_i^\psi \psibf_\ell.
\end{align*}
Compared to other similar methods in DTR or ITR estimation, dWOLS and G-dWOLS offer both double-robustness of blip estimators and relative ease in implementation given the estimating function is a weighted least squares \cite{wallace2015doubly}. Likewise, another key advantage of these estimation methods is that these approaches are regression-based. Instead of using the multi-stage framework in DTR to address longitudinal data, a simpler study design can be employed: a myopic ITR can be estimated in which repeated measurements can be accounted for by specifying a covariance structure. Combining both G-dWOLS balancing weights $w_i$ adhering to the balancing property and proposed covariance structure $V_i \in \mathbb{R}_{n_i \times n_i}$, the estimating function in which $\psibf$ needs to be solved for is similar to the one provided in step 4 of section~\ref{subsection:dwols} in which diag$(w_i)$ is multiplied by $V_i$.\b

Estimation of standard errors can be done robustly using robust or sandwich estimator (although this ignores variability due to estimation of the balancing weights) or bootstrap resampling on individuals. By performing the resampling of individuals rather than observations, the correlation structure induced by the repeated measures is preserved. However, because the number of observations varies from one individual to another, sample sizes can also differ between bootstrap iterations.
\section{Analysis of INSPIRE Data}
\label{section:data-analysis}

For this work, interest lies in  determining the smallest number of injections to ensure reconstitution of CD4 load above a healthy threshold of 500 cells$/\mu$L. In other words, we are interested in obtaining the optimal number of IL-7 injections by preventing the administration of unnecessary ones. Because the INSPIRE 2 and 3 protocol calls for injections if the CD4 load falls below 550 cells/$\mu$L at quarterly evaluations, the optimal number of injections will be determined at 90 day time intervals, which was the usual duration between two visits as per trial protocol \cite{thiebaut2014quantifying}.

%A total of 0, 1, 2 or 3 injections could be administered at the beginning of each treatment stage, all of which contain the same concentration of IL-7 at 20$\mu$g/kg \cite{thiebaut2016repeated}.

\subsection{An Outcome Balancing Benefits and Patient Inconvenience}
\label{subsection:data-adaptation}

\begin{figure*}[!htp]
  \centering
  \includegraphics[width=\textwidth]{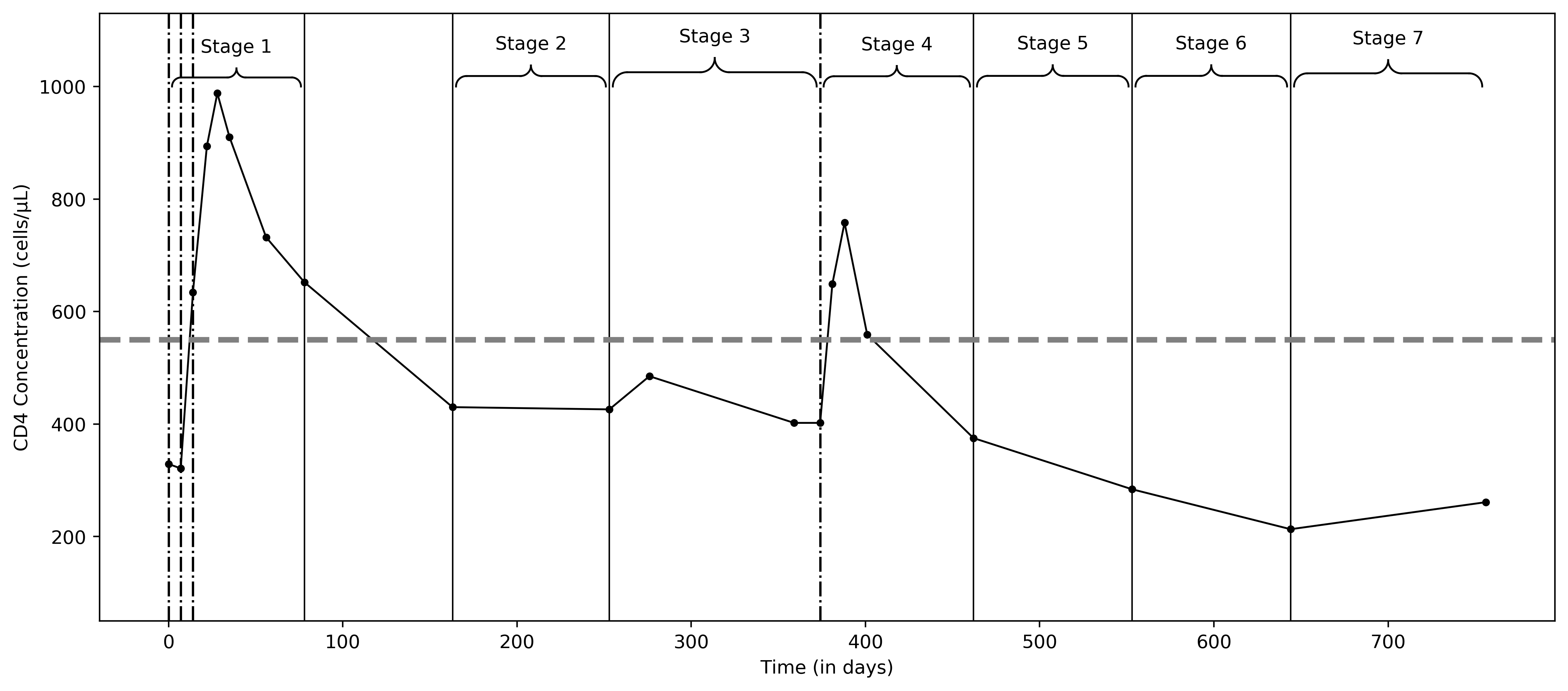}
  \caption{Estimated CD4 dynamics of an INSPIRE 2 participant using their observed CD4 counts with seven treatment stages labelled.}
  \label{fig:treatment-cycle-vs-stage}
\end{figure*}

The first challenge was to conceive a scalar-valued outcome consisting of a utility accounting for a patient's immune response and penalization for excessive injections, so that the outcome of interest was a single measurement representing an entire 90-day windows which we call a treatment stage. We also define such treatment stages to be time intervals where patients are eligible to receive injections, i.e. the first CD4 measurement of candidate treatment stages or $\text{CD4}_1$ must be below 550 cells/$\mu$L \cite{thiebaut2016repeated}. Given that participants are enrolled for at least one year,  CD4 dynamics need to be estimated. For simplicity, we define CD4$(t)$ to be the estimated trajectory of patients cell count obtained via linear interpolation (see Appendix D in the Supplementary Material for an example). More sophisticated statistical methods such as mechanistic models \cite{jarne2017modeling, prague2012treatment, villain2019adaptive} have been used to explicitly estimate the CD4 T-cell dynamics. In Figure~\ref{fig:treatment-cycle-vs-stage}, we illustrate an example of treatment stage labelling alongside an estimated CD4 trajectory for a given patient.

\begin{example}
    Consider the CD4 dynamics of a patient from INSPIRE 2 with the labelling of treatment stages in Figure~\ref{fig:treatment-cycle-vs-stage}. Here, this patient receives two sets of injections: a first set consisting of three injections for treatment stage 1 spanning from day 0 to day 374 and a second set consisting of a single injection for treatment stage 4 spanning from day 374 to day 756. Both treatment intervals are split into multiple treatment stages: the first treatment interval contains 3 treatment stages whereas the second cycle contains 4 treatment stages. In the first treatment interval, the participant receives an injection cycle consisting of 3 injections in the first treatment stage and this is followed by 3 treatment stages where no injections are administered. In the second treatment interval, the participant receives an injection cycle consisting of a single injection in the first treatment stage (4th overall) and this is followed by 3 treatment stages with no injections. Notice that the interval between day 78 and day 163 does not constitute as a treatment stage because the CD4 count measured at day 78 is above 550 cells/$\mu$L.
\end{example}

An outcome of interest denoted by $U^g$ that captures immune response is defined to be the proportion of time in a treatment stage where a participant had CD4 level above 500 cells/$\mu$L. By definition, $U^g \in [\text{0, 1}]$ and $U^g =$ 0 if all observed measurements are below 500 cells/$\mu$L. The cost $U^{\text{inj}} \in \{-\text{3}, -\text{2}, -\text{1}, -\text{0}\}$ is defined to be the negative of the number of injections. In other words, the purpose of penalizing for the number of injections on top of a low CD4 cell count is to prevent the administration of superfluous injections. Inspired by Pasin et al. (2018) \cite{pasin2018controlling}, we define the outcome variable of interest, denoted by $U(\eta)$, to be a convex sum of two previously defined utilities, $U^g$ and $U^{\text{inj}}$:
  \begin{align*}
    U(\eta) = \eta U^g + (1 - \eta) U^{\text{inj}} \quad \text{for } \eta \in [0, 1].
  \end{align*}

The parameter $\eta$ allows the focus of the constructed outcome variable to vary depending on the chosen value of $\eta$. Thus, an $\eta$ of 0 would suggest that the utility is simply the negative of the number of injections, which would be maximized by never injecting participants. Conversely, setting $\eta$ to 1 would generate a treatment rule designed to maximize CD4 response, without any consideration of the number of injections.\b

It is worth noting that the possible values that $A^\opt$ can take depends on $\eta$ due to the definition of $U$. Because $U^g$ is a proportion, i.e. $U^g \in [0, 1]$, we have that $U \geq 0$ under $A = 0$, hence the penalization term $U^{\text{inj}}$ would restrict the domain of $A^\opt$ depending on $\eta$. For instance, under $A > 0$, we have that:

\begin{align*}
\max_{U^g} U(\eta) = \eta\cdot 1 + (1 - \eta) U^{\text{inj}} \leq \eta + (1 - \eta)(-1) = 2 \eta - 1 < 0 \text{ for } \eta < 0.5 \,.
\end{align*}

Likewise, $A = $1 is only eligible to be the optimal number of injections for $\eta \geq 0.5$, $A = $2 for $\eta \geq $2/3 and $A = $3 for $\eta \geq $ 3/4 (see Appendix E for more details).

\subsubsection{Q-TWiST Method}

The results from the statistical analysis using ITR modelling will provide the ideal number of injections according to a patient's profile and response to previous injections. While we consider the potential onset of side effects important in assessing the necessity of each IL-7 injection, data related to the occurrence of adverse effects was not provided in this analysis. As such, a penalization term for the number of injections will serve as a surrogate for the risk of side effects in the statistical analysis of an optimal treatment rule.\b%From a modelling perspective, to accommodate the trade-off between immune response utility and number of injection administered, respectively represented by variable $U^g$ and $U^{\text{inj}}$ whose definitions are detailed in section~\ref{subsection:data-adaptation}, the outcome variable $U$ is defined to capture information from both utilities in which a hyperparameter $\eta$ chosen between 0 and 1 inclusively will determine the weight allocation to the two utilities. For instance, by setting $\eta =$ 0, the outcome will be the negative number of injections administered and the optimization problem would be to minimize the provision of treatment regardless of a patient's immune response. Likewise, if $\eta = 1$, all the weight would be shifted to $U^g$ and the goal would now be to maximize the duration where CD4 is above the 500 cells/$\mu$L threshold.

The idea behind the definition outcome variable $U(\cdot)$ shares similar objectives with the Q-TWiST method, short for Quality-adjusted Time Without Symptoms of disease or Toxicity of treatment \cite{gelber1991quality, gelber1995comparing}. The overarching goal behind this method is to define a single outcome which captures different utilities, all of which bears clinical relevance in evaluating the trade-off between quality and quantity of life. An outcome of interest to be investigated is formed from a weighted sum of three utilities: treatment toxicity, time spent devoid of symptoms or side effects and time after relapse \cite{gelber1995comparing}. The difference in importance allocated to factors surrounding quality or quantity of life can be reflected by varying weights values attributed to the three utilities. For instance, patients who are more prone to adverse effects may put a stronger emphasis on symptom-free time whereas others may prefer stronger treatments if it decreases the chances of illness relapse. The Q-TWiST method provides a framework to compare different treatment options for patients with potentially different clinical preferences. Particularly in the clinical management of chronic illnesses and palliative care, a quality of life index is of prime importance in assessing various treatment possibilities for late-stage diseases \cite{gelber1995comparing}. In individualizing the number of IL-7 injections, rather than considering three utilities as in the Q-TWiST method, a single outcome capturing the two quantities $U^g$ and $U^{\text{inj}}$ can be optimized under fixed values of the utility weight parameter $\eta$.

\subsubsection{Tailoring Variables}
\label{subsection:tailoring}

In practice, the idea behind longitudinal studies as a whole is to investigate the effects of studied interventions over time. In this analysis, we allow for previous immune responses to affect future ones in the modelling procedure by conditioning on individual patients' histories. That is, in addition to positing a correlation structure between treatment stages of the same patient, one way to account for historical injection information is to define tailoring variables that can embody potential prior biological response to the administration of IL-7. A patient's historical treatment information, denoted by Hx, is a dichotomous variable indicating if a patient has received injections in a prior treatment stage. A patient's response to previous treatment is denoted by Resp, and is defined as followed:
\begin{align*}
  \Resp = 
  \begin{cases}
    0\qquad &\text{if Hx } = 0\\
    \frac{1}{\# \text{ prev inj}} \Big(\underset{k}{\text{max}} \left(\left\{\text{CD4}_k^\prev \right\}_{k=1}^{n_{\text{prev}}}\right) - \text{CD4}_{1}^\prev \Big) \qquad &\text{if Hx } = 1,
  \end{cases}
\end{align*}

\noindent  where the ``prev'' superscript refers to the most recent preceding treatment stage in which injections were administered to the individual. The $k$ subscript indexes the set of observed CD4 counts in that ``prev'' treatment stage, hence $\text{CD4}^\prev_1$ refers to its first CD4 reading and $n_{\text{prev}}$ is the number of observations in that treatment stage. For instance, in Figure~\ref{fig:treatment-cycle-vs-stage}, ``prev'' refers to stage 1 for stages 2, 3, and 4 whereas, for stages 5, 6 and 7, ``prev'' refers to treatment stage 4. Resp is defined to be the largest increase in CD4 counts compared to the ``prev'' baseline value $\text{CD4}^\prev_1$ and the multiplicative coefficient $\left\{\# \text{ prev inj}\right\}^{-1}$ serves the purpose of adjusting for the number of injections administered. The idea is to highlight the sensitivity of a participants' immune response to the quantity of IL-7 provided. Since Resp exhibited considerable skew, a transformation which we denote logResp = log(Resp + 1) was employed in the G-dWOLS analysis to reduce the impact of outlying values. Data on participants' characteristics collected in the INSPIRE studies 2 and 3 can be used to tailor the number of injections to HIV-infected patients within the G-dWOLS framework. In the outcome model, such quantities can be incorporated into the blip function, and include the following information: age, sex, BMI, ethnicity and logResp. A table summarizing these variables is given in Table~\ref{table:one} in section~\ref{subsection:descriptive}.

%Because HIV-infected patients who have a CD4 load greater than $500$ cells/$\mu L$ exhibit comparable immunological benefits to a non-infected individual, the threshold of $550$ cells/$\mu L$ instead of $500$ provides a clinically sufficient buffer in.

\subsubsection{Analysis Plan}

The statistical analysis of INSPIRE data using an ITR framework is summarized in the following protocol. In what follows, the subscript $i$ indexes data points (pre-treatment covariates, number of injections, and utility) from a specific treatment stage rather than an study participant.

\begin{itemize}
  \item[1.] Compute $U^g$ and $U^{\text{inj}}$ for all patients and all stages. A value of 1 for the Origin variable indicates that a participant is of African origin and 0 otherwise (i.e.~Caucasian or other); the Sex variable was also defined as dichotomous, where 1 represents male.
  \item[2.] Fit generalized propensity scores $P(A_i = a_i \, | \, \Xbf_i^\alpha)$ where covariates $\Xbf_i^\alpha$ consist of the following: Sex, Age, BMI, Origin and $\text{CD4}_1$. Weights $w_i$ were taken to be IPT weights, i.e. $w_i = P(A_i = a_i \, | \, \Xbf_i^\alpha)^{-1}$. We use $A =$ 0 as baseline treatment as it is the largest group. % due to stronger statistical power.
  \item[3.] Define the treatment-free and blip model covariates to include patient-specific data:
  \begin{itemize}
    \item $\Xbf^\beta$: Sex, Age, BMI, Origin, Hx, logResp and $\text{CD4}_{1}$
    \item $\Xbf^\psi$: Sex, Age, BMI, Origin, Hx and logResp
  \end{itemize}
  \item[4.] Compute $U(\eta) = \eta U^g + (1-\eta)U^{\text{inj}}$ for a given weight value $\eta$ for all individual-stage data. Separate analyses will be performed for each value of $\eta$.
  \item[5.]  Apply the G-dWOLS algorithm above using the defined IPT weights $W$, treatment-free covariates $\Xbf^\beta$, blip covariates $\Xbf^\psi$ and outcome variable $U(0)$ using a weighted GEE to perform estimation (here, we opt for an independence working correlation). We use $\psi_\ell$ for the vector of blip parameters associated with injections for $a \in \{a_1, a_2, a_3\} \equiv \{$1, 2, 3$\}$ (where, recall, $A = a_0 =$ 0 is the baseline treatment category). %Minor specifications:
  \item[6.] Estimate empirical standard errors and confidence intervals using the robust variance estimator for the blip parameters $\widehat{\psibf}$ obtained from the previous step.%\footnote{Bootstrap estimation of standard errors and confidence intervals are computationally intensive but preferable as they account for the estimation of the propensity score.}.
  \item[7.] Determine the optimal treatment (number of injections) for each patient using the formula provided in step 5 of section~\ref{subsection:dwols} where $a \in \{\text{0, 1, 2, 3}\}$.
\end{itemize}

Repeat the analytic steps 4-7 for other values $\eta \in$ [0, 1].

\subsection{Results}
\label{section:results}

\subsubsection{Descriptive Results}
\label{subsection:descriptive}

Treating patient-stages as observations, a summary of its covariates is available in Table~\ref{table:one}. The standardized mean difference (SMD) is a score that measures the imbalance of characteristics across observations in different treatment groups \cite{flury1986standard, yang2012unified}. An SMD value greater than 0.1 is a common criterion used to determine if a particular covariate is imbalanced across treatment groups \cite{stuart2013prognostic}. For instance, in Table~\ref{table:one}, both the Age and Sex variables have an SMD value of 0.12, the smallest SMD value amongst covariates of interest. This descriptive measure shows that there are significant imbalances in all characteristics across observations grouped by treatment category. It is also worth highlighting the low sample size in treatment categories $A =$ 1 and $A =$ 2. There are 315 observations associated with 0 injections and 150 observations associated with 3 injections whereas there are only 17 and 22 observations associated with 1 and 2 injections respectively. The low number of observations in the $A =$ 1 and $A =$ 2 group is a by-product of the INSPIRE protocols where 3 injections should be have administered according to study protocol. The average treatment stage duration is 90.5 days with a standard deviation of 12.3 days. It is assumed that this variability does not substantially affect the analysis, especially in the definition of the immune response utility, where $U^g$ is computed as a proportion.

\bgroup
\def\arraystretch{1.3}
\begin{table}[!ht]
  \centering
  \caption[Summary of patient characteristics]{Summary of patient characteristics with respect to the number of injections received}
  \begin{tabular}[t]{cccccc}
    \toprule
    \multirow{3}{*}{Characteristic} & \multicolumn{4}{c}{Mean (SD) or Count (Proportion)} & \multirow{3}{*}{SMD}\\%changed from multirow{4}
    \\
    \cmidrule{2-5}
    & \begin{tabular}{@{}c@{}}$A=0$\\$(n=315)$\end{tabular}
    & \begin{tabular}{@{}c@{}}$A=1$\\$(n=17)$\end{tabular} 
    & \begin{tabular}{@{}c@{}}$A=2$\\$(n=22)$\end{tabular} 
    & \begin{tabular}{@{}c@{}}$A=3$\\$(n=150)$\end{tabular} 
    &\\
    \midrule
    Sex  & $227 \,\, (77\%)$ & $12 \,\, (71\%)$ & $17\,\,(77\%)$ & $102 \,\, (68\%)$ & $0.12$\\
    %$0.77\,\,(0.44)$ & $0.71\,\,(0.47)$ & $0.77\,\,(0.43)$ & $0.68\,\,(0.47)$ & $0.12$\\
    Age     & $45.4\,\,(8.8)$ & $46.3\,\,(8.8)$ & $44.5\,\,(7.9)$ & $44.9\,\,(8.4)$ & $0.12$\\
    BMI     & $24.4\,\,(3.6)$ & $25.0\,\,(4.5)$ & $25.8\,\,(4.5)$ & $24.3\,\,(3.5)$ & $0.21$\\
    Origin  & $0.40\,\,(0.49)$ & $0.29\,\,(0.47)$ & $0.32\,\,(0.48)$ & $0.48\,\,(0.50)$ & $0.22$\\
    $\text{CD4}_\init$ & $350\,\,(112)$  & $435\,\,(89)$   & $358\,\,(105)$  & $322\,\,(116)$ & $0.56$\\
    logResp    & $3.60\,\,(2.53)$  & $4.89\,\,(1.90)$  & $2.97\,\,(2.79)$  & $2.01\,\,(2.68)$ & $0.64$\\
    \bottomrule
  \end{tabular}
  \label{table:one}
\end{table}
\egroup

% \begin{figure*}[!b]
%     \centering
%     \includegraphics[width=0.8\textwidth]{example_images/U_eta_values_white.png}
%     \caption[Average $U(\eta)$ values vs. $\eta$.]{Average $U(\eta)$ values with 95\% confidence intervals with respect to treatment group plotted across $\eta$ values in [0, 1].}
%     \label{fig:outcome-plot}
% \end{figure*}

\begin{figure*}[!htp]
    \centering
    \includegraphics[width=0.8\textwidth]{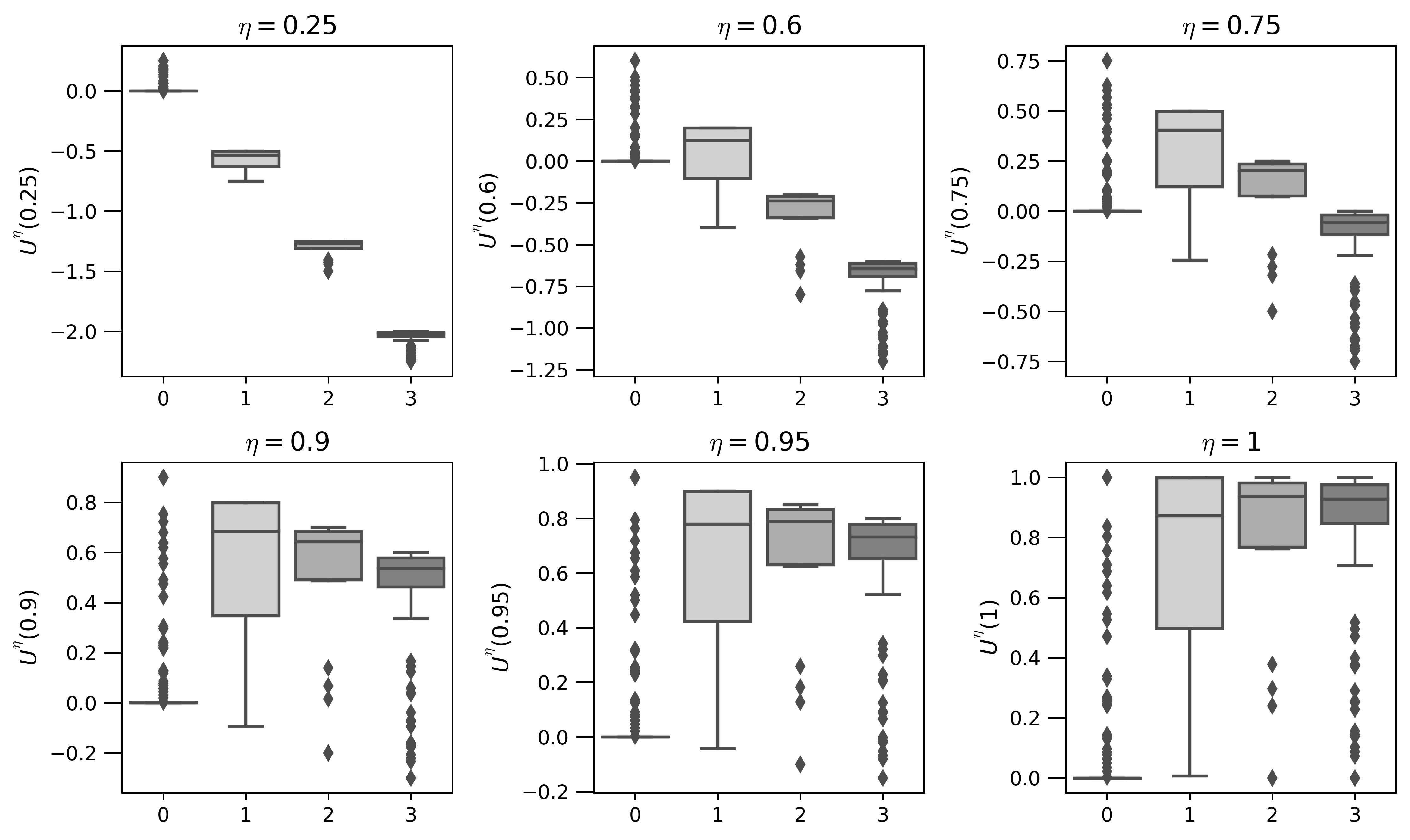}
    \caption{Boxplots for $U(\eta)$ values for $\eta \in \{\text{0.25, 0.6, 0.75, 0.9, 0.95, 1}\}$.}
    \label{fig:outcome-boxplot}
\end{figure*}

%The average $U(\cdot)$ value across observations and their 95\% confidence intervals are represented by a thick line and dotted lines, respectively. The linearity of means and confidence interval bounds follows immediately from $U(\cdot)$ being a linear combination of $U^g$ and $U^{\text{inj}}$. Boxplots for $U(\eta)$ are displayed in Figure~\ref{fig:outcome-boxplot} on the next page, stratified by the number of injections provided in patient-stages for several $\eta$ values. 
Figure~\ref{fig:outcome-boxplot} provides an overview of the outcome variable $U(\eta)$ distribution with respect to treatment categories and utility weights. For instance, for $\eta <$ 0.6, the value of $U(\cdot)$ is largely dominated by the penalizing term $U^{\text{inj}}$ for the quantity of administered injections. This result follows from the discrepancy in the widths of ranges of $U^g$ and $U^{\text{inj}}$; $U^g \in [\text{0, 1}]$ whereas $U^{\text{inj}} \in \{\text{-3, -2, -1, 0}\}$. When more emphasis is put on immune response, i.e.~in cases where $\eta$ conveys a larger value such as 0.9, 0.95 or 1, the outcome values for 1, 2 and 3 injections are closer to each other. In the bottom three plots in Figure~\ref{fig:outcome-boxplot}, utilities capturing immune response information do not seem to be considerably different in patient-stages where at least 1 injection was administered. $U^g$ values for 0 injections are 0 for nearly all patient stages, which explains the lack of variability in the boxplot for $A =$ 0.\b

Generalized propensity scores were fitted using multinomial logistic regression, which were then used to compute the weights $w_i$ needed for the dWOLS analysis. Boxplots for generalized propensity scores $P(A = a \, |\, \Xbf)$, IPT and overlap weights are displayed in Appendix F of the Supplementary Material.\b

The use of either weight functions in G-dWOLS analysis can be argued. While both IPT weights and overlap weights adhere to the balancing property for G-dWOLS, a potential gain in efficiency is possible when using overlap weights due to the restriction in their range. However, the similarity in results produced by our simulation results (see Appendix B in the Supplementary Material) using the different weight forms permits the use of either weight function in the analysis of INSPIRE studies. In fact, IPT weighting for controlling of confounders has garnered much attention from researchers in the causal inference literature over the recent year. The popularity of this well-studied and well-documented method also motivate the use of IPT weights in the data analysis of INSPIRE data \cite{austin2015moving, moodie2012q, rosenbaum1983central}.

\subsubsection{Statistical Summary of Blip Coefficients}
\label{subsection:stat-analysis}

Blip parameter estimates of $\widehat{\psibf}_{1}$, $\widehat{\psibf}_2$ and $\widehat{\psibf}_3$ and their confidence intervals are dependent on the outcome measure $U(\cdot)$ which itself relies on a particular choice of $\eta \in$ [0, 1]. Each $\eta$ value yields different values for the outcome variable, which in turn implies that the estimated ITR would vary accordingly. Because there are many coefficient estimates of interest and the utility weight is a continuous parameter, the overarching purpose of the statistical analysis and its interpretation is to paint a comprehensive portrait of the results while providing sufficient detail to determine key factors that influence treatment recommendation. The statistical analysis results for $\eta =$ 0.7 and 0.9 are provided in Tables~\ref{table:eta0.7} and~\ref{table:eta0.9} respectively on the following page. Coefficient estimates are displayed alongside their 95\% confidence intervals; bold estimates are statistically significant at a 5\% significance level. From these summary tables, coefficients for Hx and Resp, denoted by $\widehat{\psi}_{\ell, \text{Hx}}$ and $\widehat{\psi}_{\ell, \text{logResp}}$, are statistically significant across contrast functions $\gamma_\ell(\cdot), \,\ell =$ 1, 2, 3 for both $\eta =$ 0.7 and $\eta =$ 0.9. Other coefficients such as $\widehat{\psi}_{2, \text{Sex}}, \widehat{\psi}_{3, \text{Sex}}$ and $\widehat{\psi}_{3, \text{Age}}$ are also statistically significant for $\eta =$ 0.7 and $\eta =$ 0.9. For residual plots, see Appendices G and H in the Supplementary Material for outcome variables $U(0.7)$ and $U(0.9)$ respectively.
\bgroup
\def\arraystretch{1.15}
\begin{table}
  \begin{minipage}{\textwidth}
  \centering
  \caption[Summary of estimated blip coefficients with outcome $U(0.7)$.]{Summary of estimated blip coefficients for a G-dWOLS analysis of outcome $U(0.7)$}
  \label{table:eta0.7}
  \begin{tabular}{ccccccc}
    \toprule
    \multirow{2}{*}{Characteristic} & \multicolumn{6}{c}{Estimates ($95 \%$ C.I.)}\\
    \cmidrule{2-7}
    &  \multicolumn{2}{c}{$A = 1$} & \multicolumn{2}{c}{$A = 2$} & \multicolumn{2}{c}{$A = 3$}\\
    \midrule
    Intercept& $-0.11$ & $(-0.64, 0.41)$ & $-0.10$ & $(-0.83, 0.63)$ & $0.17$ & $(-0.11, 0.45)$\\
    Sex& $-0.02$ & $(-0.11, 0.06)$ & $\pmb{0.22}\footnote{\textbf{Bold} indicates statistical significance at a 0.05 significance level.}$ & $(0.00, 0.43)$ & $\pmb{-0.06}$ & $(-0.12, 0.00)$\\
    Age\footnote{\label{footnote:age}For every $10$ years}& $0.08$ & $(-0.01, 0.17)$ & $-0.07$ & $(-0.18, 0.04)$ & $\pmb{-0.04}$ & $(-0.07, 0.00)$\\
    BMI\footnote{\label{footnote:bmi}For every $10$kg/$\text{m}^2$}& $-0.03$ & $(-0.16, 0.10)$ & $0.03$ & $(-0.13, 0.19)$ & $\pmb{-0.12}$ & $(-0.21, -0.04)$\\
    Origin & $0.14$ & $(-0.02, 0.29)$ & $0.11$ & $(-0.08, 0.30)$ & $-0.02$ & $(-0.08, 0.05)$\\
    Hx& $\pmb{-1.79}$ & $(-2.69, -0.89)$ & $\pmb{-1.35}$ & $(-2.15, -0.55)$ & $\pmb{-0.65}$ & $(-1.08, -0.21)$\\
    logResp& $\pmb{0.29}$ & $(0.13, 0.45)$ & $\pmb{0.24}$ & $(0.10, 0.39)$ & $\pmb{0.10}$ & $(0.03, 0.18)$\\
    \bottomrule
  \end{tabular}
  \vspace{3mm}
  \caption[Summary of estimated blip coefficients with outcome $U(0.9)$.]{Summary of estimated blip coefficients for a G-dWOLS analysis of outcome $U(0.9)$}
  \label{table:eta0.9}
  \begin{tabular}{ccccccc}
    \toprule
    \multirow{2}{*}{Characteristic} & \multicolumn{6}{c}{Estimates ($95 \%$ C.I.)}\\
    \cmidrule{2-7}
    &  \multicolumn{2}{c}{$A = 1$} & \multicolumn{2}{c}{$A = 2$} & \multicolumn{2}{c}{$A = 3$}\\
    \midrule
    Intercept& $0.14$ & $(-0.54, -0.82)$ & $0.44$ & $(-0.50, 1.4)$ & $\pmb{1.08}$ & $(0.72, 1.44)$\\
    Sex& $-0.03$ & $(-0.14, 0.08)$ & $\pmb{0.28}$ & $(0.04, 0.56)$& $\pmb{-0.08}$ & $(-0.20, 0.00)$\\
    $\text{Age}^{\text{\ref{footnote:age}}}$& $0.10$ & $(-0.01, 0.21)$ & $-0.09$ & $(-0.23, 0.06)$ & $\pmb{-0.05}$ & $(-0.10, 0.00)$\\
    $\text{BMI}^{\text{\ref{footnote:bmi}}}$& $-0.04$ & $(-0.20, 0.13)$ & $0.04$ & $(-0.17, 0.25)$ & $\pmb{-0.16}$ & $(-0.28, -0.05)$\\
    Origin& $0.18$ & $(-0.03, 0.38)$ & $0.14$ & $(-0.11, 0.38)$ & $-0.02$ & $(-0.11, 0.06)$\\
    Hx & $\pmb{-2.30}$ & $(-3.45, -1.14)$ & $\pmb{-1.73}$ & $(-2.77, -0.70)$ & $\pmb{-0.83}$ & $(-1.39, -0.27)$\\
    logResp & $\pmb{0.37}$ & $(0.17, 0.57)$ & $\pmb{0.31}$ & $(0.12, 0.50)$ & $\pmb{0.13}$ & $(0.033, 0.23)$\\
    \bottomrule
  \end{tabular}
  \end{minipage}
\end{table}
\egroup

\begin{figure*}[!htp]
    \centering
    \includegraphics[width=\textwidth]{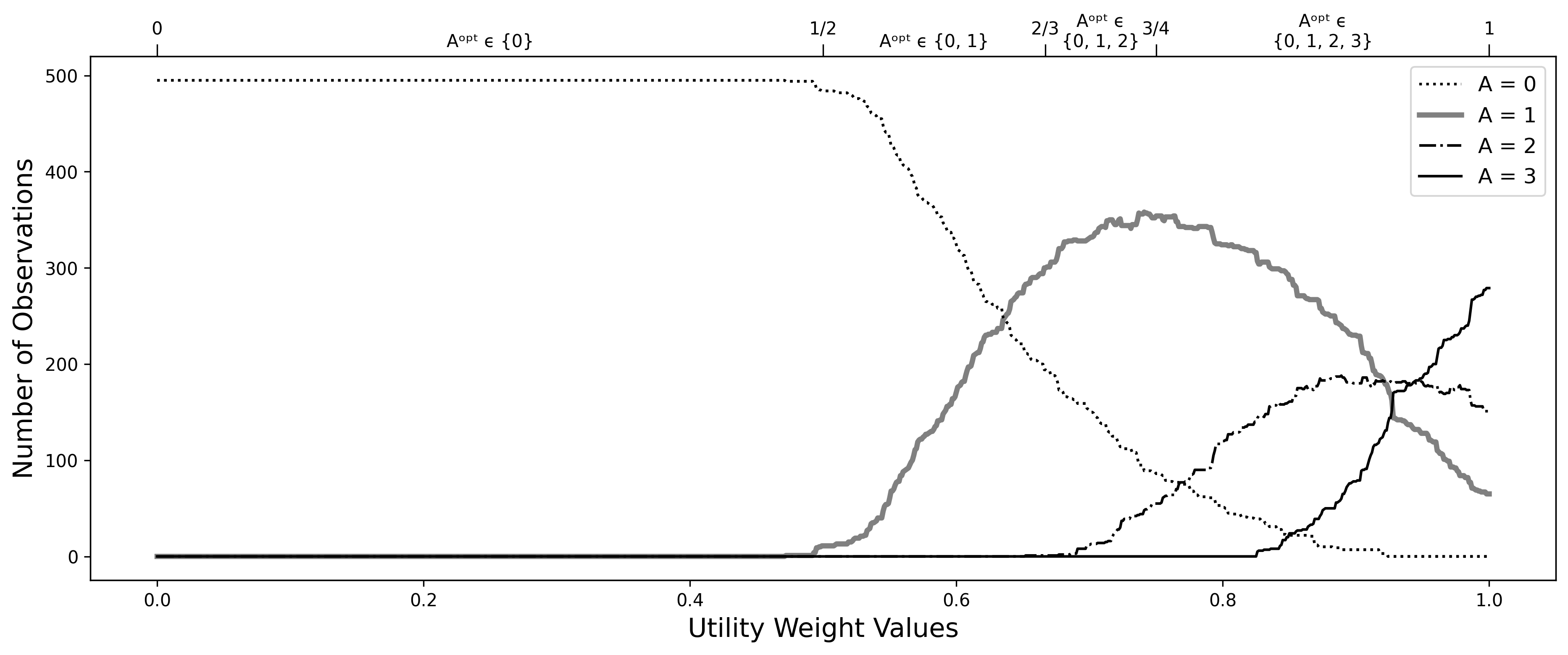}
    \caption[Number of observations having $A^{\text{opt}}$ as each treatment type.]{Number of observations having $A^{\text{opt}}$ as each treatment type with respect to $\eta$ values. Possible values of $A^\opt$ as a function of $\eta$ are indicated on the top horizontal axis.}
    \label{fig:optimal-injections}
\end{figure*}

The optimal number of injections can also be calculated for each patient-stage in the dataset used for the analysis. Estimation of contrast values $\gamma_\ell(\xbf^\psi)$ can be done by plugging in covariate-specific information $\xbf^\psi$ for each patient-stage in the processed dataset. $\widehat{A}^\opt$ can be obtained by comparing values of the blip functions as discussed in section~\ref{subsection:dwols}. The number of observations having $\widehat{A}^\opt$ being equal to 0, 1, 2 or 3 injections are displayed with respect to $\eta$ values ranging from 0 to 1 in Figure~\ref{fig:optimal-injections}. One important thing to observe is that, when $\eta =$ 1, i.e.~when there is no penalization for the number of injections, a considerable number of patient-stages are still being recommended 1 or 2 injections rather than 3. Although other factors such as a patient's age, sex, ethnic origin may affect the estimated ITR, recall that the observed immune response utility $U^g$ seems to be comparable across groups $A =$ 1, $A =$ 2 and $A =$ 3 (see Figure~\ref{fig:outcome-boxplot}).

\subsubsection{Treatment Recommendation for Specific Patient Profiles}

\begin{figure*}[!ht]
  \centering
  \begin{minipage}{.45\textwidth}
    \centering
    \includegraphics[width=\linewidth]{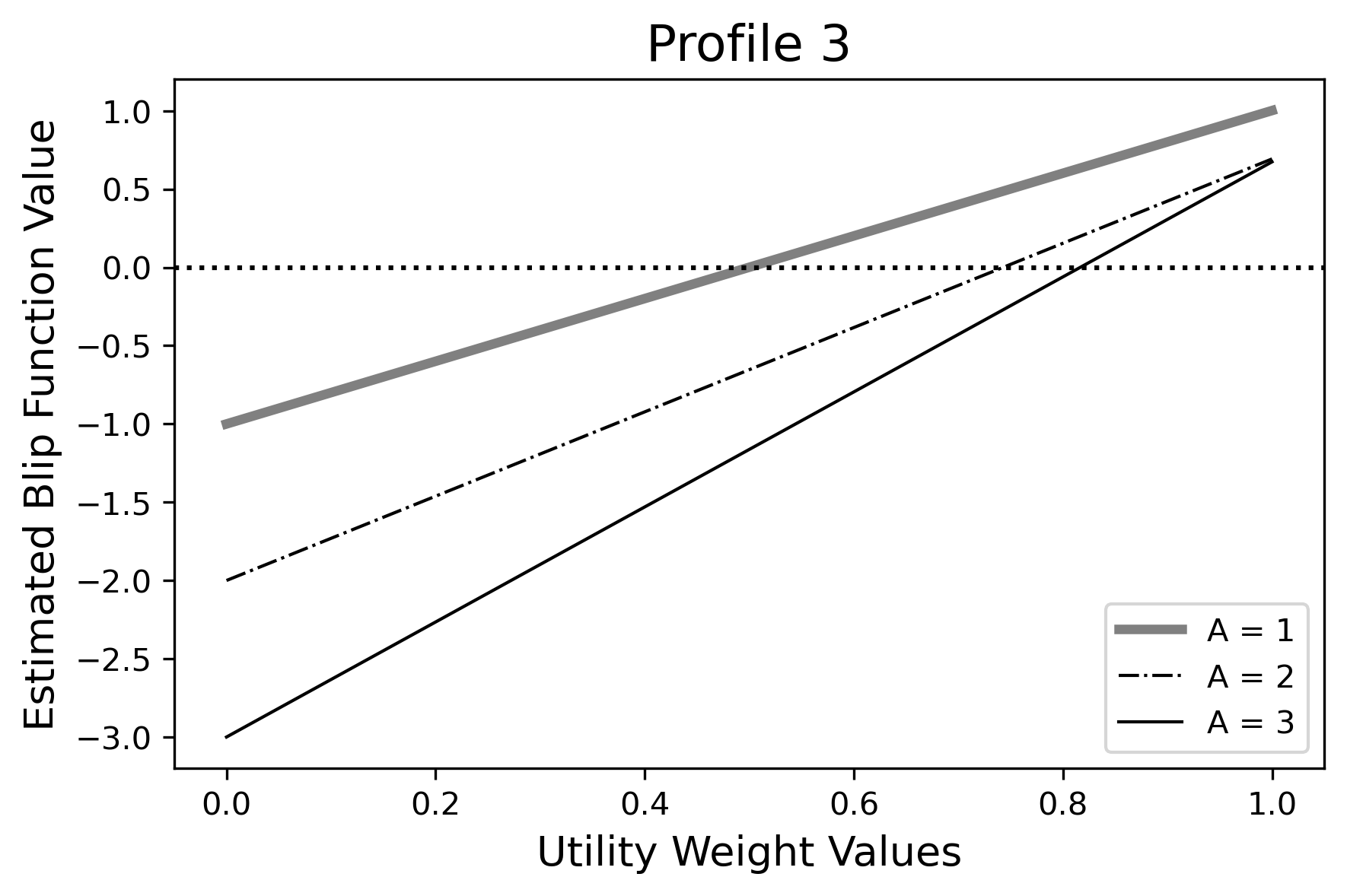}
  \end{minipage}
  \begin{minipage}{.45\textwidth}
    \centering
    \includegraphics[width=\linewidth]{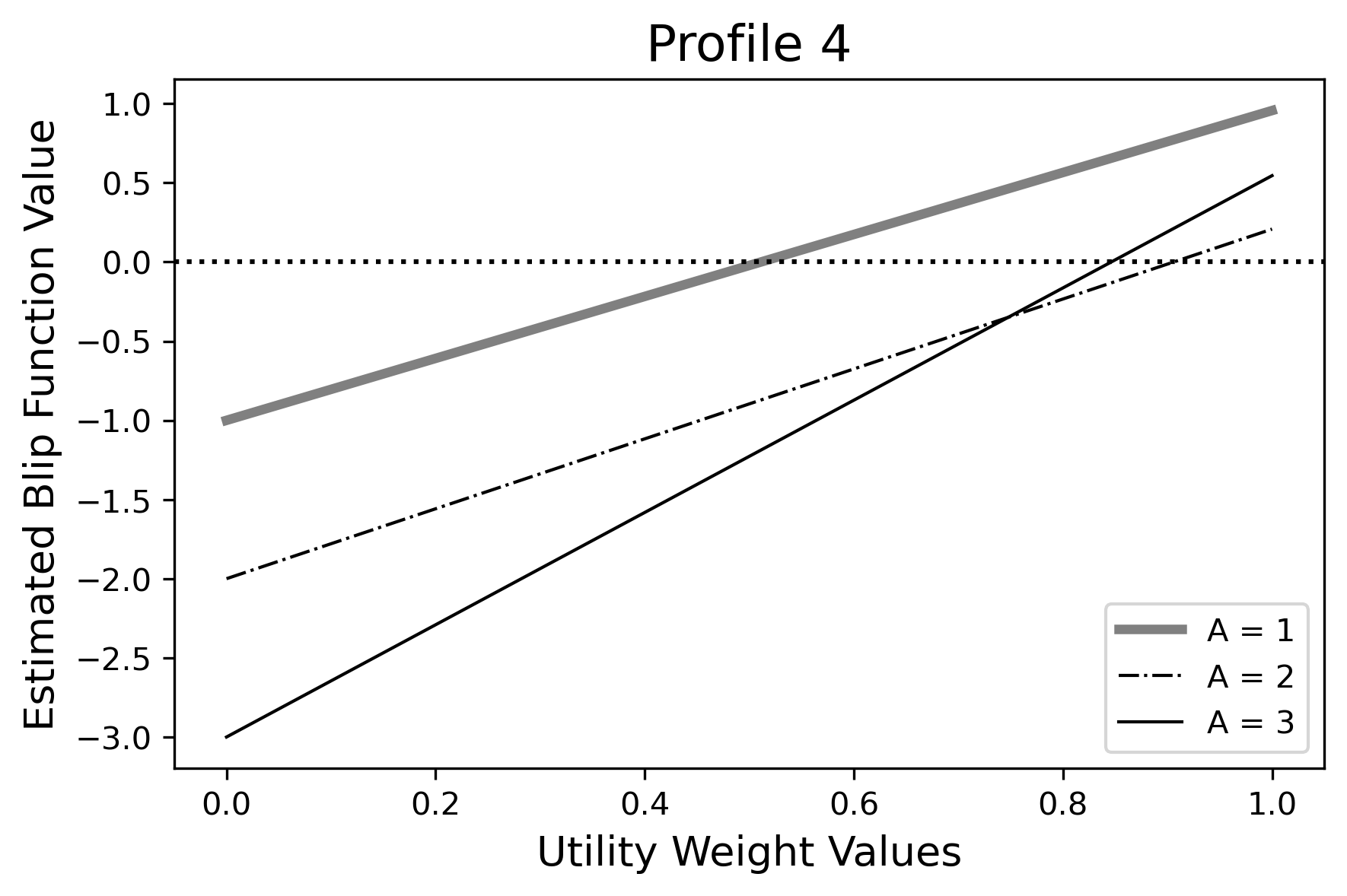}
  \end{minipage}
  \caption[Patient profile contrast function utility values.]{Contrast function utility for different number of injections with respect to utility weights $\eta \in [0, 1]$ for four patient profiles.}
  \label{fig:contrast}
\end{figure*}

Treatments depend on the chosen weight values associated with utilities $U^g$ and $U^{\text{inj}}$; however the individualization of the treatment recommendation also depends on characteristics of patients, as the blip model is a function of covariates $\Xbf^\psi$. To give an idea of the optimal number of injections, plots showing the estimated value of $\gamma_\ell(\cdot, a_\ell)$ for $a_\ell =$ 1, 2, 3 for four chosen patient profiles are shown in Figure~\ref{fig:contrast}. The chosen patient profiles are as follow:
\b
\textbf{Profile 1}: 25-year-old man of non-African ethnic background, with a BMI of 25 and without any IL-7 injection history;\\[1mm]
\textbf{Profile 2}: 40-year-old woman of African ethnic background, with a BMI of 35 and with a recent injection cycle with a Resp value of 200;\\[1mm]
\textbf{Profile 3}: 60-year-old man of African ethnic background, with a BMI of 24 and without any IL-7 injection history;\\[1mm]
\textbf{Profile 4}: 80-year-old woman of non-African ethnic background, with a BMI of 30 and with a recent injection cycle with a Resp value of 400.
\begin{figure*}[!ht]
  \centering
  \begin{minipage}{.45\textwidth}
    \centering
    \includegraphics[width=\linewidth]{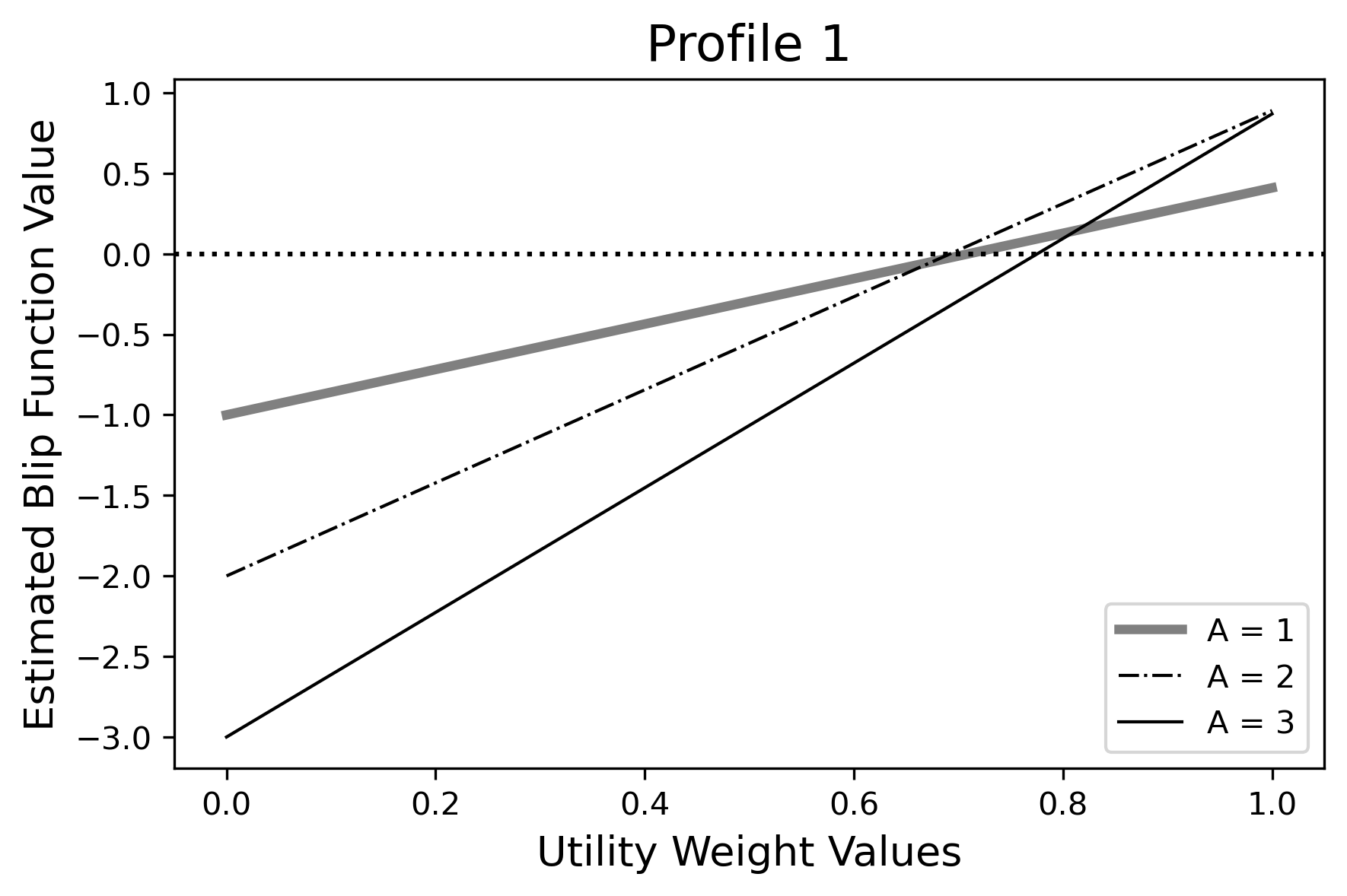}
  \end{minipage}
  \begin{minipage}{.45\textwidth}
    \centering
    \includegraphics[width=\linewidth]{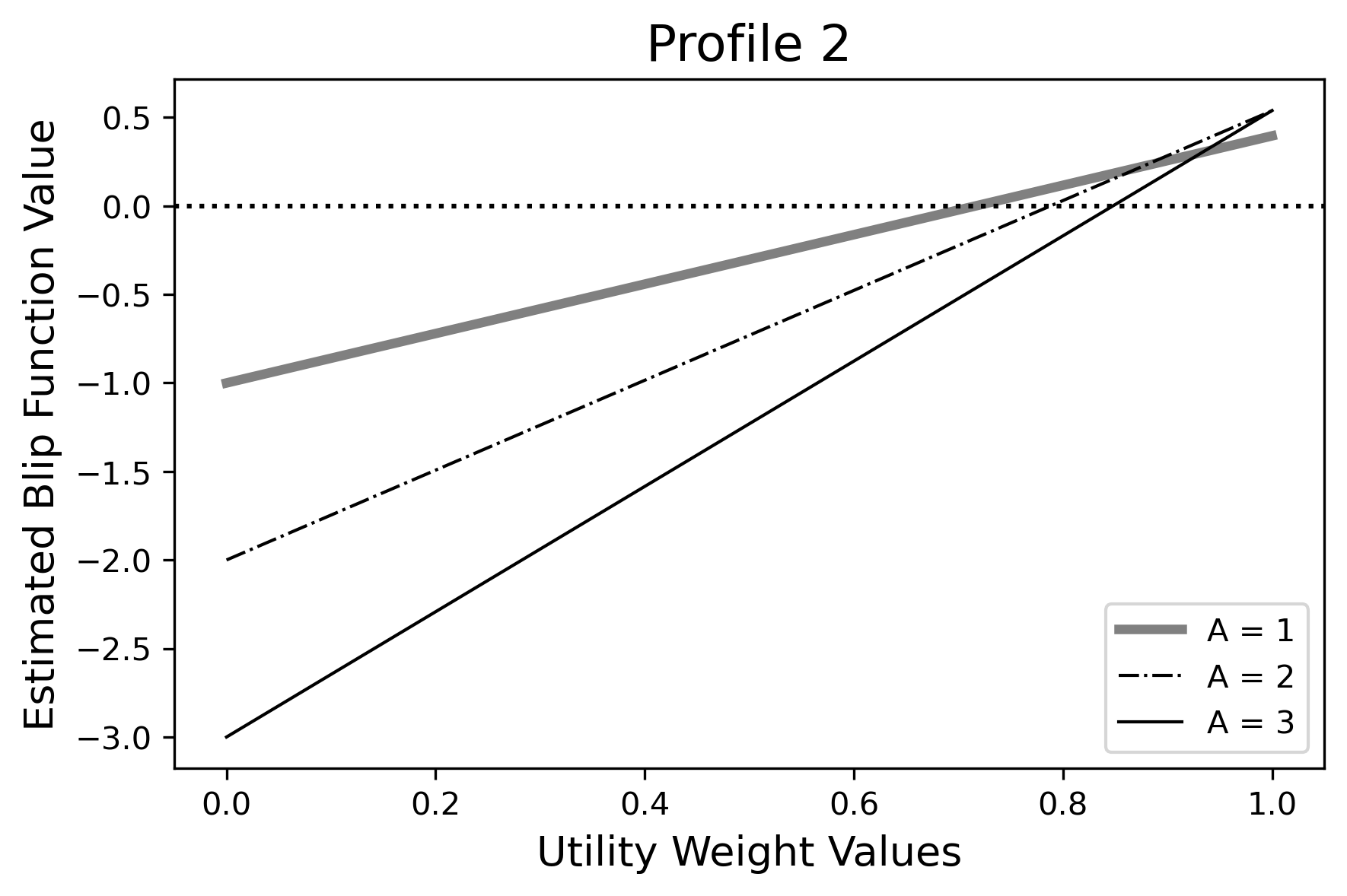}
  \end{minipage}
\end{figure*}
\FloatBarrier
The dotted line representing the 0 utility threshold is the value of the blip function for the $A =$ 0 category since it was chosen to be reference treatment. Because the blip function compares utilities across treatment options, the number of injections exhibiting the largest utility value for a fixed $\eta$ value is the optimal treatment determined by the dWOLS analysis. Profile 1 seems to suggest that $A = 2$ is generally better than other treatment option counterparts while treatments seem to offer similar estimated utility values in Profile 2 (with $A = 2$ marginally better than $A = 1, 3$ for $\eta > 0.9$. Estimated contrast function plots for Profiles 3 and 4 show that a single injection seems to perform better than two and three injections for $\eta > 0.5$. For values of $\eta < 0.5$, no injections seem to be the recommended treatment for all four patient profiles. These results align well with the theoretical range of $A^\opt$ as discussed in Section~\ref{subsection:data-adaptation} and in Appendix E. From these plots, three conclusions can be drawn.\b

Firstly, similarly to what \cite{jarne2017modeling} have found, there does not seem to be a considerable benefit in administering 3 injections. Instead, either 1 or 2 injections seem to result in comparable or even better immune response since, for $\eta =$ 1, $A = 2$ seems to provide better utility than $A  = 3$ for patient profiles 1 and 2 whereas $A = 1$ curves are above $A = 3$ for patient profiles 3 and 4. Secondly, although $A^\opt = 0$ for $\eta < 0.5$, the decision to administer a non-zero number of injections seems to vary across patient profiles. The $A = 1$ line crosses the dotted line in profile 3 and 4 for a smaller value of $\eta$ relative to the $A = 2$ line in profiles 1 and 2. The main difference between these pairs of patient profiles is their age: profiles 1 and 2 are respectively 20 and 40 years old whereas profiles 3 and 4 are respectively 60 and 80 years old. This in turn can imply that, according to this analysis, a first IL-7 injection is more necessary in older participants than in younger counterparts but a second dosage is unnecessary. However, for younger patients, assigning 1, 2 or 3 injections yield comparable results in immune response; a moderate decision of 2 injections may be the best, most conservative approach.
\section{Discussion and Conclusion}

The goal of this statistical analysis was to show that G-dWOLS can be applied on longitudinal data within a GEE framework in constructing a myopic ITR. Overall, the G-dWOLS analysis provides valuable insight on the ideal number of injections through the design of ITR, one for each value of the utility weight $\eta$. Treatment recommendation largely depends on two things: the value of $\eta$ and patient-specific information. When $\eta$ is closer to 0, more weight is put on minimizing the number of injections and, as a result, the recommended number of injections is conservative. When $\eta$ is closer to 1, more importance is put on having a better immune response, hence participants are more likely to be recommended to receive injections. However, despite this, many observations in the dataset are suggested to receive 1 or 2 injections. This result follows immediately from the good immune response outcome for observations associated with these treatment categories, despite their low sample size (see Figure~\ref{fig:outcome-boxplot}). For a relatively large number of treatment stages, the maximal number of injections set by the clinical protocol are not being recommended; this  warrants further investigation on the relevance of a third injection first; this question was previously raised by Jarne et al. (2017) \cite{jarne2017modeling}. While a third injection may increase the CD4 load to a higher peak, their results show that two injections are sufficient to ensure that the CD4 concentration exceeds  the threshold of 500 cells/$\mu$L \cite{jarne2017modeling, pasin2018controlling}.\b

Previous approaches to optimizing IL-7 treatment protocols have not incorporated covariate-specific information on the ideal number of injections to provide \cite{pasin2018controlling, villain2019adaptive}. In this work, intrinsic patient characteristics such as BMI and ethnic origin have not shown to be either statistically or clinically significant in individualizing the number of injections. Their the estimated coefficients did not significantly differ from 0, and do not seem to alter treatment recommendation (see Tables~\ref{table:eta0.7} and~\ref{table:eta0.9}). The coefficient estimates for sex across treatment groups were also low in magnitude, except for $A =$ 2; however, this finding was not found to have a strong impact on our results. Although our analyses were limited in power, our results suggest that age may be clinically useful tailoring variable to determine the optimal number of injections. For instance, the findings in the analysis of specific patient profiles recommend a first injection in older patients at lower utility weights than in younger patients. However, for the latter subpopulation, a second injection seems more necessary whereas, in older patients, treatment recommendation seems to follow a ``one or nothing'' approach to injections. This could potentially point to the risk of treatment toxicity in older people living with HIV, since the penalization term $U^{\text{inj}}$ in this analysis served as a proxy for any potential undesirable consequences attributed to IL-7 injections, or simply to a plateauing of response in older patients. The effect of age on treatment recommendation warrants further investigation, particularly as adverse effects to IL-7 were not available to us for this analysis.\b

Two major limitations that must be raised are the sparsity of observations associated with 1 and 2 injections and the design of the myopic rule. When receiving a cycle of IL-7 injections, participants from the INSPIRE studies were more much more likely to receive 3 injections than they are to receive 1 or 2; cycles consisting of 1 or 2 injections are considered incomplete by the clinical protocols of INSPIRE 2 and 3 \cite{thiebaut2016repeated, villain2019adaptive}. An inherent limitation of this is that we cannot exclude confounding by indication due to participants not receiving all 3 injections, although we consider the possibility unlikely due to the nature of the treatment and outcome. Second, the design of the studied ITR assumes that the effect of IL-7 injections are short-term, in that their lingering influence on the CD4 count across subsequent stages are insignificant. The use of an ITR to investigate injection effects is motivated by the desire to maximize immediate utilities and its simpler interpretation \cite{petersen2007individualized}. However, an observable trend in the INSPIRE participants' CD4 dynamics is its gradual decrease over time following injections. An important model-based consideration in future work would to incorporate the duration since prior injections because, with increasing time, patients CD4 load are be more likely to fall below 500 cells$/\mu$L. Although we attempted to capture information regarding immune response from previous injections, the temporal aspect was not considered in this analysis.\b

The initial research question that inspired this study addresses the issue of an ideal number of IL-7 injections to provide to a patient exhibiting poor restoration of CD4 cells despite receiving HAART. The application of G-dWOLS requires the definition of an outcome variable that takes into account information regarding both a desirable immune response and the number of injections for an entire treatment stage. In our formulation of the statistical methodology, the outcome is defined as a sum of utilities which reflect both pieces of information through a weighting parameter $\eta$ that varies between 0 and 1. As such, the optimal number of injections can only be determined for fixed values of this utility weight $\eta$. The choice of this parameter can depend on many aspects surrounding clinical outcomes of IL-7 administration, such as improvements in quality of life, treatment fatigue, side effects and treatment toxicity amongst many others. However, an ideal value of $\eta$ cannot be determined by simply examining patient characteristics and medical history: it is a value that is subjective to a patient's preference or a medical decision maker's opinion. As argued in literature surrounding the Q-TWiST method, a single clinical measure accounting for multiple health factors provides a useful framework for medical decision-making \cite{gelber1995comparing, glasziou1998quality}. The ability to evaluate treatments by shifting focus or changing utility weight values with respect to clinical preference enables decision makers and especially patients to subjectively tailor medical care. The risk-benefit assessment of the Q-TWiST approach suggests that the optimal utility weight value $\eta$ depends on the clinical implications of the statistical analysis and preferences of a patient \cite{gelber1995comparing}. In fact, it may even be the case that these preferences change with time. We posit that it is not possible to provide an optimal $\eta$ value via data-driven methods. Rather, our analysis allows patients to determine optimal number of injections with respect to their personal (rather than statistical) valuation of the potential risks and benefits of IL-7 therapy, which may change over time as, for example, side effects are experienced and their severity and tolerability assessed.\b

In this work, we have provided the first ITR analysis of the `dosing' of IL-7 using a myopic implementation of G-dWOLS applied to longitudinal data. Our analysis relied on several assumptions, including that of no unmeasured confounding which is plausible in the INSPIRE setting. The decision of an individual patient not to follow trial protocol and receive all three planned injections is highly unlikely to be associated with immune response. There have been some investigations into the use of random effects models to address unmeasured confounding \cite{papadogeorgou2019adjusting}, an avenue for future investigation in a repeated measures setting such as this. Another important direction for further research is to more fully determine the necessary and sufficient conditions under which the myopic approach may be used in place of a DTR analysis.

\pagebreak
\bibliographystyle{acm}
\bibliography{main}

\pagebreak
\section*{Appendix A: INSPIRE Protocol}
\label{appendix:a.1}

Patients assigned to the control arm of INSPIRE 2 only have CD4 and Ki67 evaluations in their ``Induction'' phase as they do not receive IL-7 injections until a year of follow-up.

\begin{figure}[!htp]
    \centering
    \includegraphics[width=130mm]{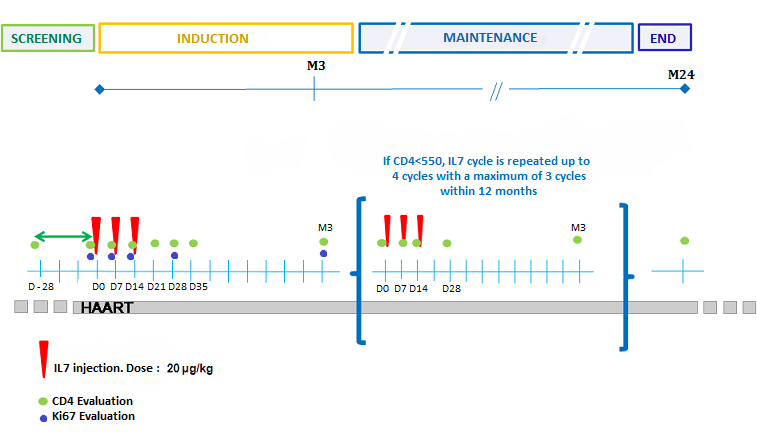}
    \caption{Protocols for INSPIRE 2 and IL-7 arm in INSPIRE 3 (\cite{jarne2017modeling, villain2019adaptive}). Three IL-7 injections, each at dose of 20$\mu$g/Kg, were provided within the first 3 weeks following enrolment into the study. After 3 months, study participants received another cycle of three injections if their CD4 count fell below 550 cells/$\mu$L. CD4 counts were measured at various times points and are displayed by green dots. Although measurement of the Ki67 markers was also taken and depicted by blue dots in the figure above, they were not relevant in our analysis.}
\end{figure}
\FloatBarrier

\section*{Appendix B: Simulation Study}
\label{appendix:a.2}

We carry out a simulation study to showcase the theoretical guarantees of the G-dWOLS analysis in devising a myopic treatment regime using longitudinal data. The goal of this simulation study is twofold; firstly, as presented in \cite{wallace2015doubly}, we want to illustrate the double robustness of blip parameter estimators by conducting simulations with varying sample sizes under different specifications of the treatment-free and treatment model. Secondly, we show that the myopic formulation of the optimization problem is appropriate for longitudinal data under the assumption of no delayed treatment effects. We generate longitudinal data to mimic the evolving nature of the outcome variable and time-varying confounders on which the G-dWOLS algorithm can be applied. Intra-patient correlation due to the contribution of multiple observations to the processed data is also created in the simulated data. The presentation of simulation results is done in two parts: a summary of blip parameter estimates under different simulation designs is first depicted followed by an application of the simulated results on a ``test'' population to better understand the potential advantages of the G-dWOLS' theoretical guarantees from a practical perspective.  That is, we examine bias and variability of the parameters that together specify the estimated optimal ITR, and then evaluate the impact of that ITR on a hypothetical population of new patients.% more ``practical'' approach in understanding the theoretical guarantees of G-dWOLS.

\subsection*{Data Generation Procedure}

We let $\mathcal{O}_{ij} = \left(\text{Sex}_{ij} \equiv \text{Sex}_i, \text{CD4}_{ij}, A_{ij}, Y_{ij}\right)$ denote the $j$th stage-specific data of individual $i$ as we generate four observations, each of which we refer as a stage, for each participant. We define $Y_j$ to be some clinical outcome to be maximized and $A_j \in \{\text{0, 1, 2}\}$ to be a categorical exposure or treatment variable. Two covariates are defined: $\text{Sex}_{i}$, a time-invariant binary variable which denote a person's biological sex, and $\text{CD4}_{ij}$, a continuous time-varying variable which can represent a person's evolving CD4 cell count. Each individual is associated with four observations $\mathcal{O}_{\cdot j}$, as if these measurements were taken at each stage in a longitudinal study. The Sex variable is sampled from a Bernoulli distribution with $p =$ 0.7 whereas the CD4 variable is sampled from a truncated normal distribution denoted by $\text{TRN}_{[50, 550]}\left(350, 100\right)$ (centered around 350 with standard deviation 100 whose domain is bounded by 50 from below and 550 from above). The distribution parameters were chosen either be representative of ``real life'' or ideal in illustrating the theoretical properties of G-dWOLS; treatment effects and qualitative visualization of simulation results can vary depending on the specified model parameters. For instance, when parameter estimates are theoretically biased, the variation in CD4 counts needs to be somewhat substantial such that the bias is visually evident. The proportion of males in participants from INSPIRE studies 2 and 3 is around 70\%, hence the $p$ was specified to be 0.7 in the distribution of the Sex variable. The upper bound value of the truncated normal distribution was selected for context completeness because, in the INSPIRE protocol, patients are only eligible for treatment if their measured CD4 load is below 550 cells$/\mu$L. On the other hand, CD4 count is a strictly positive quantity and a lower bound of 50 cells$/\mu$L was chosen to reflect a lower limit of detection for testing. Treatment allocation probabilities $p_0$, $p_1$ and $p_2$ depend on parameters $\alphabf_1$ and $\alphabf_2$, which were chosen such that a reasonable proportion of generated ``people'' are assigned to each treatment category. The outputs from Algorithm~\ref{simulation} will serve as inputs for the G-dWOLS estimation procedure which was provided in more detail in section~\ref{subsection:dwols}. The assigned stage-specific treatment is sampled from a multinomial distribution whereby the probabilities $p_0,$ $p_1$ and $p_2$ for treatment allocation are defined such that they can be estimated using a logistic regression model:
\begin{align*}
    p_0 &= \frac{1}{1 + \Xbf^\alpha \alphabf_1 + \Xbf^\alpha \alphabf_2}\text{ , and}\\
    p_{\ell} &= \frac{\Xbf^\alpha \alphabf_{\ell}}{1 + \Xbf^\alpha \alphabf_1 + \Xbf^\alpha \alphabf_2} \quad \text{for } \ell = 1, 2 \, .
\end{align*}
where $X_{ij}^\alpha = \left(1, \text{Sex}_{ij}, \text{CD4}_{ij}\right)$. Although the probabilities of receiving assigned treatment $a_0,$ $a_1$ or $a_2$ are in fact covariate dependent, the indices $i$ and $j$ are dropped to alleviate the notation. We define our outcome variable $Y_{ij}$ as followed:

\begin{align*}
    Y_{ij} = X_{ij}^\beta \betabf + \one_{A_{ij} = 1}X_{ij}^\psi \psibf_1 + \one_{A_{ij} = 2}X_{ij}^\psi \psibf_2 + b_i + \epsilon_{ij}
\end{align*}

\noindent where $X_{ij}^\beta = \left(1, \exp\left\{ \text{CD4}_{ij}/200\right\}, \sqrt{\text{CD4}_{ij}} \right)$ and $X_{ij}^\psi = \left(1, \text{Sex}_{ij}, \text{CD4}_{ij}\right)$. Intra-subject correlation is generated by a time-invariant random effect $b_i$ sampled from a normal distribution centered at zero with standard deviation 0.5 and a random effect $\epsilon_{ij}$ was sampled from a $\mathcal{N}(0, 3)$ distribution. In a first simulation study, parameters were chosen to be: $\alphabf_1 =$ (-0.5, -0.2, 0.005), $\alphabf_2 =$ (-1, -0.4, 0.007), $\betabf =$ (45, -10, 1), $\psibf_1$ = (-10, 5, 0.02) and $\psibf_2 =$ (-30, -7, 0.1). The simulations are also conducted under the null, where there is no difference in expected utilities across all treatment categories. In other words, a second set of simulations is also performed with $\psibf_1 = \psibf_2 =$ (0, 0, 0) and same values for $\alphabf_1, \alphabf_2$ and $\betabf$ as in the first simulation. A summary of the data generation procedure is available in Algorithm~\ref{simulation}.

\RestyleAlgo{boxruled}
\LinesNumbered
\begin{algorithm}[H]
    \KwIn{Parameters $\alphabf, \betabf, \psibf$ and sample size $n$}
    \caption{Data Generation Procedure for Simulation Study}
    \label{simulation}
    \For{i in \{1, \dots, n\}}{
        Sample $b_i \sim \mathcal{N}(0, 0.5)$\\
        \For{j in \{1, \dots, 4\}}{
            \eIf{j == 1}{
                Sample $\text{Sex}_{ij} \sim \text{Bern}(p)$\\
                Sample $\text{CD4}_{ij} \sim \text{TRN}_{[50, 550]}(350, 100)$
            }{
                Set $\text{Sex}_{ij} = \text{Sex}_{i1}$\\
                Sample $\Delta \sim \mathcal{N}(0, 5)$\\
                Set $\text{CD4}_{ij} = \text{CD4}_{i(j-1)} + \Delta$\\
                \uIf{$\text{CD4}_{ij} > 550$}{
                    Set $\text{CD4}_{ij} = 550$
                }\uElseIf{$\text{CD4}_{ij} < 50$}{
                    Set $\text{CD4}_{ij} = 50$
                }
            }
            Sample $A_{ij} \sim \text{Multinomial}(p_0, p_1, p_2)$\\
            % Define $X_{ij} = \left( \text{Sex}_{ij}, \text{CD4}_{ij}\right)$\\
            Define $X_{ij}^\alpha = \left(1, \text{Sex}_{ij}, \text{CD4}_{ij}\right)$\\
            Define $X_{ij}^\beta = \left(1, 
            \exp\left\{ \text{CD4}_{ij}/200\right\},
            \sqrt{\text{CD4}_{ij}}
            \right)$\\
            Define $X_{ij}^\psi = \left(1, \text{Sex}_{ij}, \text{CD4}_{ij}\right)$\\
            Sample $\epsilon_{ij} \sim \mathcal{N}(0, 3)$\\
            Set $Y_{ij} = X_{ij}^\beta \betabf + \one_{A_{ij} = 1}X_{ij}^\psi \psibf_1 + \one_{A_{ij} = 2}X_{ij}^\psi \psibf_2 + b_i + \epsilon_{ij}$
        }
    }
    \tcc{Generating inputs for G-dWOLS estimation process of simulated data}
    % Define $\Xbf^\alpha = \left \{X_{ij}^\alpha \right \}_{(i, j) \in \{1, \dots, n\} \times \{1, \dots, K\} }$\\
    % Define $\Xbf^\beta = \left \{X_{ij}^\beta \right \}_{(i, j) \in \{1, \dots, n\} \times \{1, \dots, K\} }$\\
    % Define $\Xbf^\psi = \left \{X_{ij}^\psi \right \}_{(i, j) \in \{1, \dots, n\} \times \{1, \dots, K\} }$\\
    Define $\textbf{Sex} = \left\{\text{Sex}_{ij} \right\}_{(i, j) \in \{1, \dots, n\} \times \{1, \dots, K\} }$\\
    Define $\textbf{CD4} = \left\{\text{CD4}_{ij} \right\}_{(i, j) \in \{1, \dots, n\} \times \{1, \dots, K\} }$\\
    Define $\Abf = \left \{A_{ij} \right \}_{(i, j) \in \{1, \dots, n\} \times \{1, \dots, K\} }$\\
    Define $\Ybf = \left \{Y_{ij} \right \}_{(i, j) \in \{1, \dots, n\} \times \{1, \dots, K\} }$\\
    \KwRet{$\textbf{Sex}, \textbf{CD4}, \Abf, \Ybf$}
\end{algorithm}
\FloatBarrier

\subsection*{Estimation Under Various Model Specifications}

While parameters for the generalized propensity score, the treatment-free and blip models are all estimated in the G-dWOLS algorithm, interest predominantly lies in the blip coefficient estimates $\psibf$ as they provide information on the expected difference in utilities compared to baseline treatment $A =$ 0. The specification of the set of covariates on which outcome regression and estimation of generalized propensity score are performed will dictate the asymptotic behaviour of $\widehat{\psibf}$ estimates. Recall that, in G-dWOLS, the double robustness property allows misspecification of at most one of the nuisance models to ensure consistency of blip parameter estimators.
\begin{itemize}
    \item Model 1 (both treatment model and treatment-free model incorrectly specified):
    \begin{itemize}
        \item Regress $Y$ on 1, Sex, CD4 and treatments, with interactions between treatments and each of the two covariates.
        \item Fit an intercept only multinomial logistic regression\footnote{An intercept only multinomial logistic regression model does not estimate $P(A = a \, | \, \Xbf^\alpha)$ for each value of $\Xbf^\alpha$ but rather a marginal or unconditional probability, i.e.~$P(A = a \, | \, \Xbf^\alpha) = P(A = a) = n^{-1}\big(\sum_{i=1}^n \one_{A_i = a} \big)$.}.
    \end{itemize}
    \item Model 2 (correct treatment model, incorrect treatment-free model):
    \begin{itemize}
        \item Regress $Y$ on 1, Sex, CD4 and treatments, with interactions between treatments and each of the two covariates.
        \item Fit a multinomial logistic regression on 1, Sex, CD4.
    \end{itemize}
    \item Model 3 (incorrect treatment model, correct treatment-free model):
    \begin{itemize}
        \item Regress $Y$ on 1, $\exp\left\{\text{CD4/200}\right\}$, $\sqrt{\text{CD4}}$ and treatments, with interactions between treatments and each of the two covariate, where the interaction between treatment and CD4 is on the untransformed, linear CD4 scale.
        \item Fit an intercept only multinomial logistic regression.
    \end{itemize}
    \item Model 4 (both models correctly specified):
    \begin{itemize}
        \item Regress $Y$ on 1, $\exp\left\{\text{CD4/200}\right\}$, $\sqrt{\text{CD4}}$ and treatments, with interactions between treatments and each of the two covariates, where the interaction between treatment and CD4 is on the untransformed, linear CD4 scale.
        \item Fit a multinomial logistic regression on 1, Sex, CD4.
    \end{itemize}
\end{itemize}
Even though correlation was included in the data generation procedure to induce bias caused by repeated measurements, an independence correlation structure will be used in the G-dWOLS estimation process within the GEE framework. A correct specification of the covariance structure would yield more efficient estimators, but one of the main benefits of the GEE framework in accommodating longitudinal data is its asymptotic unbiasedness of linear estimates (\cite{diggle2002analysis}). The standard error estimation can be done via using the sandwich estimator or bootstrapping, but we used the former method to alleviate computational burden.

The simulation study was conducted on sample sizes $n =$ 100 and $n =$ 1000 to investigate sample properties of estimators. As mentioned in section~\ref{subsection:dwols}, IPT and overlap weights both satisfy the balancing property, a feature of the dWOLS algorithm that ensures the double robustness of $\widehat{\psibf}$; both weights were used in analyzing the simulated data. Each simulation design requires a choice of sample size $n$, G-dWOLS weight function (IPT or overlap) and one of four possible model specifications. To numerically assess the underlying properties of our statistical methodology, 1000 Monte Carlo replications were carried out for all possible simulation scenario combinations.

\subsection*{Simulation Results}
\label{subsection:simulation-results}

The presentation of the results from the simulation study will be split into two parts: a summary of blip coefficient estimates followed by an application of these estimates on a completely novel or ``test'' set of 10 000 subject-level data. In a first instance, an overview of the variation and median of the $\widehat{\psibf}$ estimated values across the $1000$ replications would provide a better idea on the effects of sample sizes, weight function choice and model specification on the blip coefficient estimates. Afterwards, the application of simulation results on an out-of-sample population would shed light on the quantitative benefits of the G-dWOLS estimation procedure in tailoring treatment to individual-level information.\\

\begin{figure*}[!htp]
    \centering
    \includegraphics[width=\textwidth]{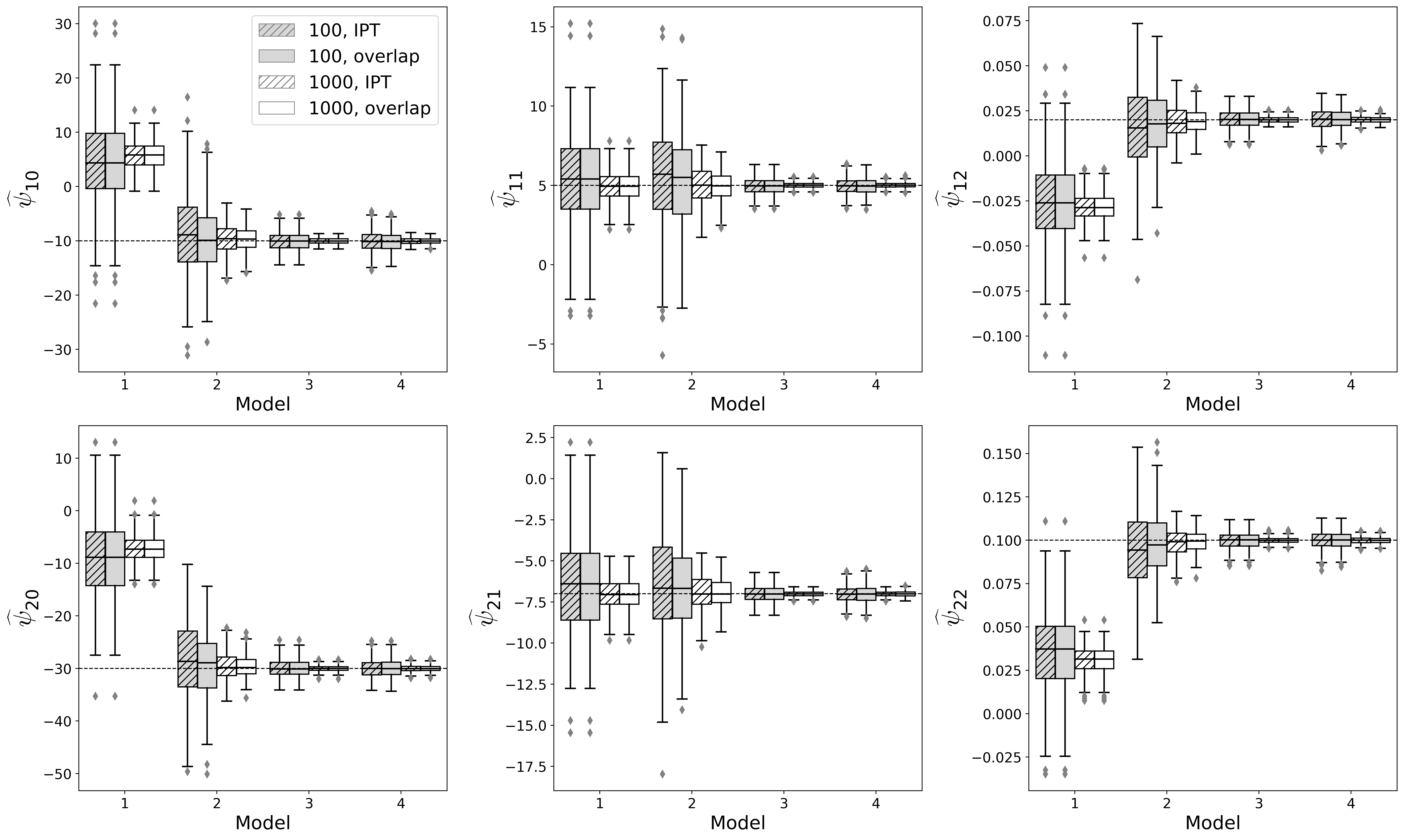}
    \caption[Boxplots of $\widehat{\psibf}$ estimates under different simulation designs.]{Boxplots of $\widehat{\psibf}$ estimates are displayed for four different model specifications, for two different sample sizes ($n =$ 100, 1000) and for two forms of balancing weights (IPT and overlap). The top row corresponds to the blip parameters for treatment level 1 (relative to level 0) and the bottom row to treatment level 2. The first column represents the main effect of treatment level, the middle and right columns are, respectively, the interactions of treatment with Sex and CD4.}
    \label{fig:simulation-1}
    \includegraphics[width=\textwidth]{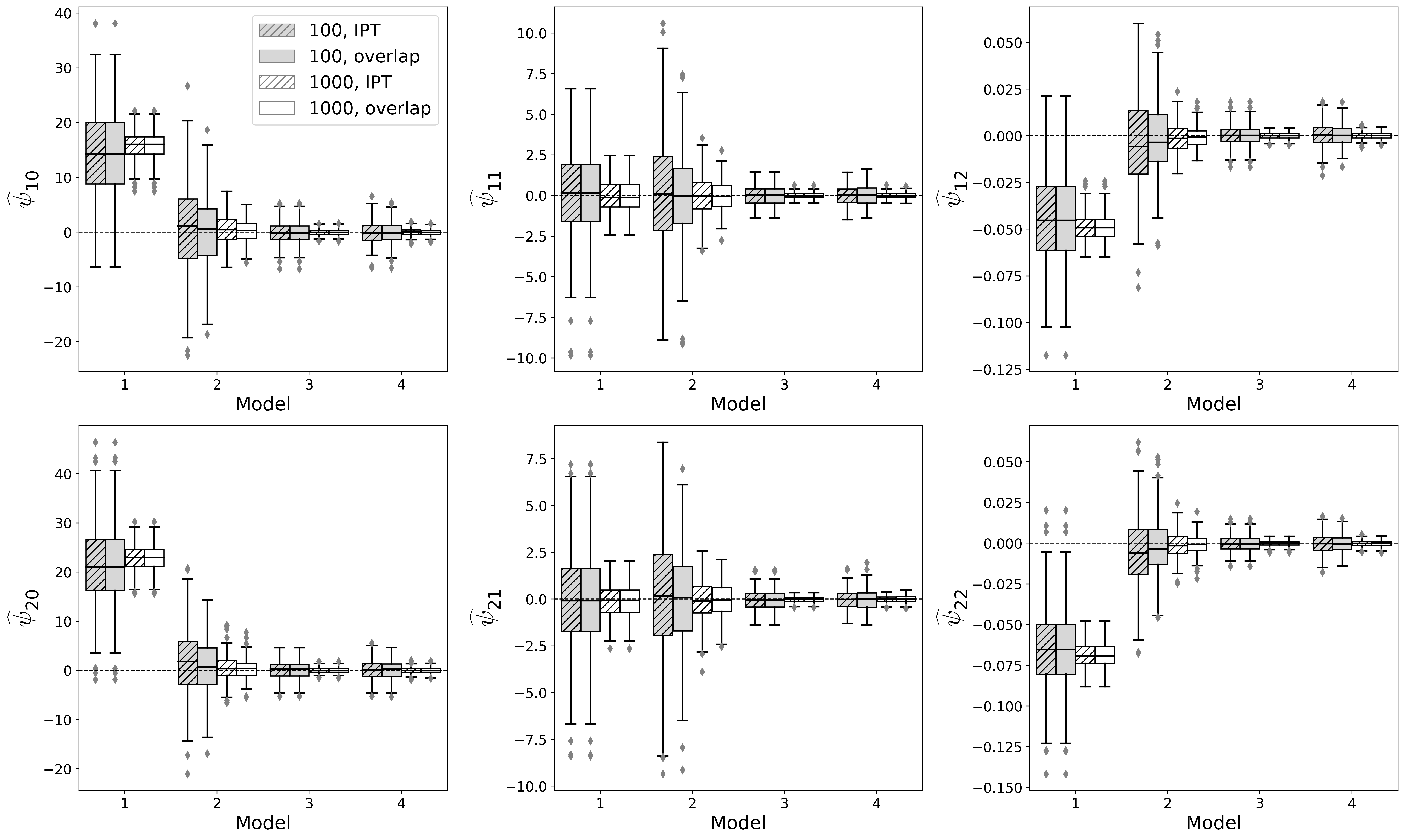}
    \caption[Boxplots of $\widehat{\psibf}$ estimates under different simulation designs and no treatment effect.]{Boxplots of $\widehat{\psibf}$ estimates are displayed for four different model specifications, for two different sample sizes and for two forms of balancing weights whereby all blip parameters set to 0 in the data generating process. Rows and columns are as in Figure~\ref{fig:simulation-1}.}
    \label{fig:simulation-null}
\end{figure*}

\textbf{Summary of Blip Coefficient Estimates} -- Each point estimate of $\psibf$ is obtained by applying the G-dWOLS algorithm on a generated dataset consisting of $n$ subjects, for $n =$ 100 or $n =$ 1000 as mentioned previously. The results of the simulation study are showcased in Figures~\ref{fig:simulation-1} and~\ref{fig:simulation-null}, which display boxplots summarizing the estimates across the 1000 runs for the non-null and null simulation settings. For purposes of notational simplicity, $\psi_{\cdot 1}$ is the blip model coefficient for Sex whereas $\psi_{\cdot 2}$ is that for CD4, with the $\cdot$ representing treatment level 1 or 2, relative to baseline level of treatment. Because similar conclusions can be drawn from the numerical results presented in both figures, the focus of this discussion will be on Figure~\ref{fig:simulation-1}.

The first row is parameters corresponding to treatment $A =$ 1 whereas the second row corresponds to treatment $A =$ 2. In each plot, there are boxplots grouped with respect to the four different combinations of model specifications as described on the previous page (Models 1, 2, 3 and 4). Labelled by their shading and hatch pattern, each set itself showcases the simulation results under different sample sizes and G-dWOLS weight form. The dotted horizontal line in every plot indicates the true value of the parameter used in the data generating process. The most important visual pattern across the plots is the bias of estimated values under Model 1. This result is not surprising, as G-dWOLS does not guarantee consistency when both nuisance models are incorrectly specified. However, it is important to point out that the estimates of $\psi_{11}$ and $\psi_{21}$ under Model 1 do not depart significantly from the true parameter value. In other words, unbiasedness can still be achieved but it is not guaranteed in Model 1. Other noticeable patterns are the decreasing interquartile range from $n =$ 100 to $n =$ 1000, which is evidence of a gain in efficiency with increasing sample size. The relative closeness between the median of estimates in Models 2, 3 and 4 and the true parameter values depicted by horizontal dotted lines suggests that unbiasedness can be reached in reasonable sample sizes. Although the simulation results show that finite sample performance improves with increasing values of $n$, $\psibf$ estimates at $n =$ 100 perform well and data analysis results using similar sample sizes can yield good point estimates with reasonable variability provided that researchers are confident in the modelling choices. The difference in using IPT weights and overlap weights, which can be inferred by comparing boxplots with diagonal hatch patterns to the ones without such patterns, does not seem to be dramatic. That being said, the smaller variability in $\widehat{\psibf}$ estimates when using overlap weights suggest that having them bounded between 0 and 1 provide a more 
precise point estimate, since IPT weights are unbounded from above (\cite{wallace2015doubly}).\\% This conclusion is indicative of little difference in the analysis between the two possible G-dWOLS weight specifications. As such, the data analysis in section~\ref{section:results} can be carried out with either IPT weights or overlap weights; IPT weights were primarily for convenience purposes.

\begin{figure}[!htp]
    \centering
    \includegraphics[width=\textwidth]{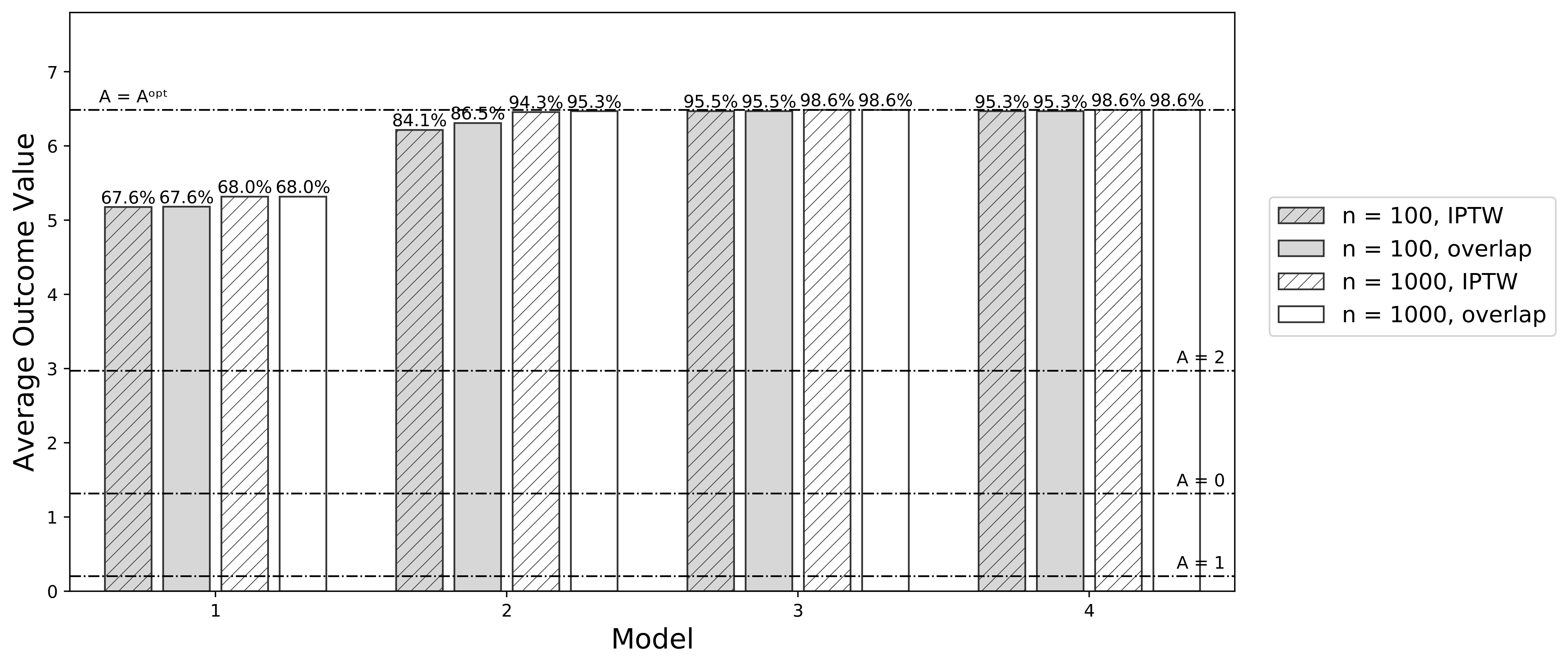}
    \caption[Average outcome under different simulation designs.]{Average outcome under different model specifications using blip coefficients estimated with different sample sizes; the corresponding agreement rate is displayed on top of each bar. Dotted lines are the average utilities under uniform treatment, labelled respectively by the treatment options $A =$ 0, 1 or 2; average utility under optimal treatment is indicated by $A^{\opt}$.}
    \label{fig:simulation-2}
\end{figure}

\textbf{Average Utilities and Agreement Rates} -- A collection of data comprised of out-of-sample 10 000 subjects, which have not been used in the estimating process, is generated using Algorithm~\ref{simulation} as well to further examine and validate the results of the simulation study. Since this additional dataset was not used in obtaining estimates of blip function parameters, two comparisons which stem from this newly created set of potential participants are made. Firstly, the expected outcome value under uniform treatment (e.g.~all 10 000 ``test'' people receive medical intervention $A =$ 0, 1 or 2) are evaluated. Secondly, the utility under the estimated optimal treatment denoted by $\widehat{A}^{\opt}$ depends on the value of $\widehat{\psibf}$ of a given iteration. The estimated ideal treatment $\widehat{A}^\opt$ in patient-stages will be compared to the true values of $A^\opt$, whereby the proportion of $A^\opt$ agreeing with $\widehat{A}^\opt$ will be referred to as the agreement rate. In Figure~\ref{fig:simulation-2}, we show the average outcome value under $\widehat{A}^{\opt}$ for the out-of-sample data which we define as $\overline{Y}^{\opt} = \sum_{i=1}^{10\,000} \E[Y \, | \, \Xbf=\xbf_i,\, A = \widehat{A}^{\opt}; \widehat{\psibf}]$. Note that the true values of $A^\opt$ are known for all patient-stages since blip coefficient parameters are known by design in a simulation study, fixed by the choice of data-generating parameters.

In the absence of substantial variability in the outcome variable or treatment effects, a perfect agreement rate can be attained in drawing meaningful conclusions regarding the effect of sample size, forms of G-dWOLS weight functions and model specifications on the estimation of optimal treatment. The values of $\overline{Y}^{\opt}$ are represented by the height of each bar in Figure~\ref{fig:simulation-2} for various model specifications. The corresponding agreement rate for each simulation design is displayed on top of the bars in the barplot. Results confirm that the selection of weight function for the G-dWOLS algorithm provides comparable results, although the overlap weights seem slightly better in Model 2 where only the treatment is correctly specified. Models 2, 3 and 4 show that larger outcome values and better agreement rates can be achieved in light of the correct specification of at least one of the treatment-free or treatment model. However, for $n =$ 100, Model 3 and 4 seem to outperform Model 2, which suggests that the gain in efficiency can be slightly different in smaller sample sizes depending on which nuisance model is correctly specified. Nonetheless, the main conclusion that can be extracted from Figures~\ref{fig:simulation-1} and~\ref{fig:simulation-2} is that desirable results are achievable in relatively larger sample sizes given that at least one of the nuisance models is correctly specified and the chosen weight function adheres to the balancing property.

\section*{Appendix C: Suitability of the Myopic Strategy for the INSPIRE Setting}

Suppose that $Y = Y_1 + Y_2$, i.e.~the final, end-of-study outcome of interest in the dynamic treatment strategy is the sum of the stage-specific outcomes, $Y_j$, and consider the regret functions for a two stage setting:
\begin{eqnarray*}
\mu_1(\xbf_1, a_1; \psibf) &=& \E[Y(a_1^{opt},a_2^{opt}(a_1^{opt})) - Y(a_1,a_2^{opt}(a_1)) |  \Xbf_1 = \xbf_1, A_1 = a_1] \\
\mu_2(\xbf_2, a_2; \psibf) &=& \E[Y(a_1,a_2^{opt}(a_1)) - Y(a_1,a_2) |  \Xbf_1 = \xbf_1, A_1 = a_1, \Xbf_2 = \xbf_2, A_2 = a_2 ].
\end{eqnarray*}

\noindent We begin with the second-stage optimal treatment. The optimal \textit{dynamic} (rather than myopic) treatment strategy maximizes $Y$, and hence minimizes the regret:
$$a_{d,2}^{opt} = \arg\max_{a^* \in \mathcal{A}_2} \E[Y(a_1, a^*) - Y(a_1,a_2) |  \Xbf_1 = \xbf_1, A_1 = a_1, \Xbf_2 = \xbf_2, A_2 = a_2 ]. $$
As we condition on all pre-treatment covariates and the first stage treatment, note that $Y_1$ must be fixed and cannot be affected by choice of the second-stage treatment. Thus the counterfactual first-stage outcome, $Y_1(a_1,a_2)$, is a function of $a_1$ only. That is, we have that
\begin{eqnarray*}
Y(a_1,a_2) &=& Y_1(a_1,a_2)  + Y_2(a_1,a_2) \\
           &=& Y_1(a_1)  + Y_2(a_1,a_2).
\end{eqnarray*}
Hence,
\begin{eqnarray*} && \arg\max_{a^* \in \mathcal{A}_2} \E[Y(a_1, a^*)) - Y(a_1,a_2) |  \Xbf_1 = \xbf_1, A_1 = a_1, \Xbf_2 = \xbf_2, A_2 = a_2 ] \\
&& \qquad = \arg\max_{a^* \in \mathcal{A}_2} \E[Y(a_1, a^*)) |  \Xbf_1 = \xbf_1, A_1 = a_1, \Xbf_2 = \xbf_2 ] \\
&& \qquad = \arg\max_{a^* \in \mathcal{A}_2} \E[Y_2(a_1, a^*)) |  \Xbf_1 = \xbf_1, A_1 = a_1, \Xbf_2 = \xbf_2 ] \\
&& \qquad = \arg\max_{a^* \in \mathcal{A}_2} \E[Y_2(a_1, a^*)) - Y_2(a_1,a_2) |  \Xbf_1 = \xbf_1, A_1 = a_1, \Xbf_2 = \xbf_2, A_2 = a_2 ].
\end{eqnarray*}
This is sufficient to show that the optimal myopic treatment strategy at the second stage coincides with the optimal dynamic strategy at this stage, as the first expression defines the optimal dynamic strategy and the latter defines the optimal myopic strategy.

Let us turn now to the optimal dynamic treatment strategy in the first stage, defined by
\begin{eqnarray*} &&\arg\max_{a^* \in \mathcal{A}_1} \E[Y(a^*,a_2^{opt}(a^*)) - Y(a_1,a_2^{opt}(a_1)) |  \Xbf_1 = \xbf_1, A_1 = a_1] \\
&& \qquad = \arg\max_{a^* \in \mathcal{A}_1} \E[Y(a^*,a_2^{opt}(a^*))  |  \Xbf_1 = \xbf_1] \\
&& \qquad = \arg\max_{a^* \in \mathcal{A}_1} \E[Y_1(a^*) + Y_2(a^*,a_2^{opt}(a^*)) |  \Xbf_1 = \xbf_1].
\end{eqnarray*}

The optimal myopic treatment strategy at the first stage maximizes $Y_1(a^*)$ over $a^* \in \mathcal{A}_1$; denote this $a_1^{opt,m}$ where the $m$ indicates `myopic'. If $a_1^{opt,m}$ is the maximizer of $Y_2(a^*,a_2^{opt}(a^*))$ over $a^* \in \mathcal{A}_1$, then clearly the myopic and dynamic treatment strategies at this first stage will also coincide. In what circumstances, then, does this hold? A sufficient set of conditions for the two strategies to coincide are (1) treatment effects are immediate, such that $A_j$ impact $Y_k$ for $k=j$ only, and (2) there are no synergistic or antagonistic effects between treatments at different stages. The latter condition implies that there are no statistical interactions between treatments, and hence that prior treatments do not act as tailoring variables. In the INSPIRE study, these sufficient conditions are biologically plausible because, when including lagged injections as a tailoring variable in the blip function on top of covariates detailed in Section 4.3, their statistical coefficients are not statistically significant at a 5\% level for selected values of $\eta$, i.e. $\eta = $0.7 and 0.9 as we used in our analyses.

% \textcolor{red}{LARRY: can you please double check this in two ways. First, double check with Rodolphe or in the lit that IL-7 is unlikely to have long-term or carry-over effects (so we can say that it won't affect future $Y$s \textbf{and} unlikely to interact with later injections in any way. Second, could you do some really simple analyses to check if our outcomes as we define them ($U$s) seem to depend on \textbf{lagged} injections and/or there is any evidence of an interaction between current and lagged injection -- just looking one interval back is sufficient. Any results you can add on this (e.g.~even just stating that the lagged effects do not significantly interact with current number of injects and do not significantly predict outcome would be helpful.}

\section*{Appendix D: Example of Linear Interpolation}

Here, we provide an illustrative example on how we construct our outcome variable $U^g$ in the adaptation of the INSPIRE data for the analysis of ITR.

\begin{example}
  Consider a patient's CD4 dynamics in Figure~\ref{fig:linearinterpolation} in which they have three observations. Two injections were provided in this treatment stage and they are represented by the two vertical dotted lines. The three CD4 measurements are 400, 700 and 500 taken at respectively day 0, 24 and 80.
  \begin{figure*}[!ht]
    \centering
    \includegraphics[width=0.9\textwidth]{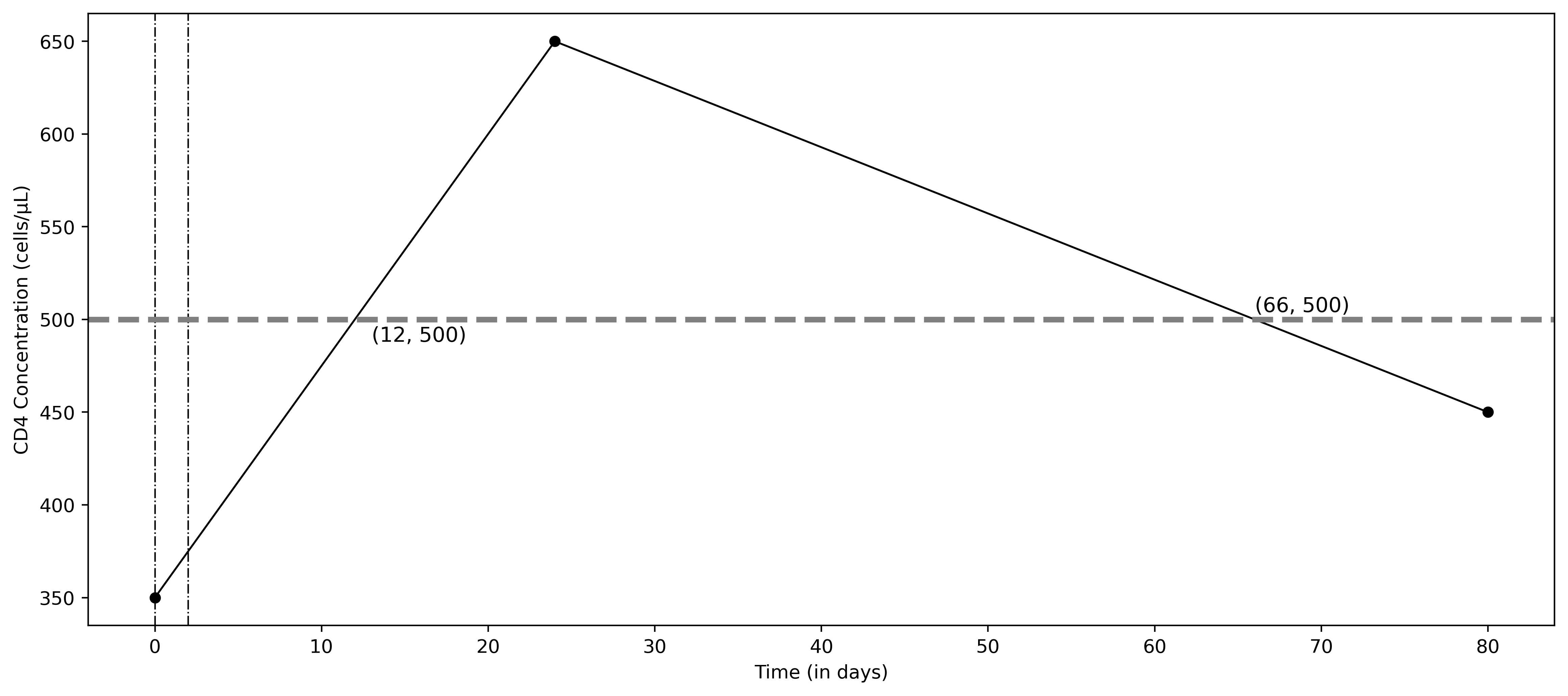}
    \caption{Estimation of CD4 dynamics using linear interpolation.}
    \label{fig:linearinterpolation}
  \end{figure*}
  \FloatBarrier
  Because two injections were administered, $U^i =$ -2. The estimated CD4 trajectory in Figure~\ref{fig:linearinterpolation} crosses the 500 cells/$\mu$L threshold at time points $t \in \{\text{12, 66}\}$. The utility associated to immune response can be calculated as follows:
    \begin{align*}
    U^g = \frac{1}{80}\int_{0}^{80} \one_{\text{CD4}(t) \geq 500} \,\text{d}t = \frac{1}{80} \int_{12}^{66} 1 \,\text{d}t = 0.55
  \end{align*}
\end{example}

\section*{Appendix E: Range of $A^\opt$ with respect to $\eta \in [0, 1]$}

Recall that our outcome is defined as $U(\eta) = \eta U^g + (1 - \eta)U^{\text{inj}}$, where $U^g \in [0, 1]$ and $U^{\text{inj}} \in \{0, -1, -2, -3\}$. We can evaluate the range of $A^\opt$ by comparing $U(\eta)$ under $A = k$ for $k = 1, 2, 3$ to $0$, the minimal value of $U(\eta)$ under $A = 0$. In other words, because the utility is a non-negative value if no injections are administered, $A = k$ cannot be preferred for certain values of $\eta$ due to the penalization of $U^{\text{inj}}$.

\begin{align*}
    \max_{U^g} \left\{\eta U^g + (1 - \eta)U^{{\text{inj}}}\right\} &\leq 0 \qquad \text{under regime } A = k\\
    1 + (1 - \eta)(-k) &\leq 0\\
    \eta &\leq \frac{k}{1 + k}
\end{align*}

This implies that, for $k = 1$, $U(\eta) \leq 0 \Leftrightarrow A^\opt \neq 1$ for $\eta \leq \frac{1}{2}$. Likewise, the same logic can be applied for $k = 2$ and $k = 3$. As a result, we have that:

\begin{align*}
    A^\opt &\in \{0\} \quad \text{for } \eta \in \Big[0, \frac{1}{2}\Big]\,;\\
    A^\opt &\in \{0, 1\} \quad \text{for } \eta \in \Big(1/2, 2/3\Big]\,;\\
    A^\opt &\in \{0, 1, 2\} \quad \text{for } \eta \in \Big(2/3, 3/4\Big]\,;\\
    A^\opt &\in \{0, 1, 2, 3\} \quad \text{for } \eta \in \Big(3/4, 1\Big]\, .\\
\end{align*}

\section*{Appendix F: Boxplot of Generalized Propensity Scores and Weights}

\begin{figure*}[!b]
  \includegraphics[width=\textwidth]{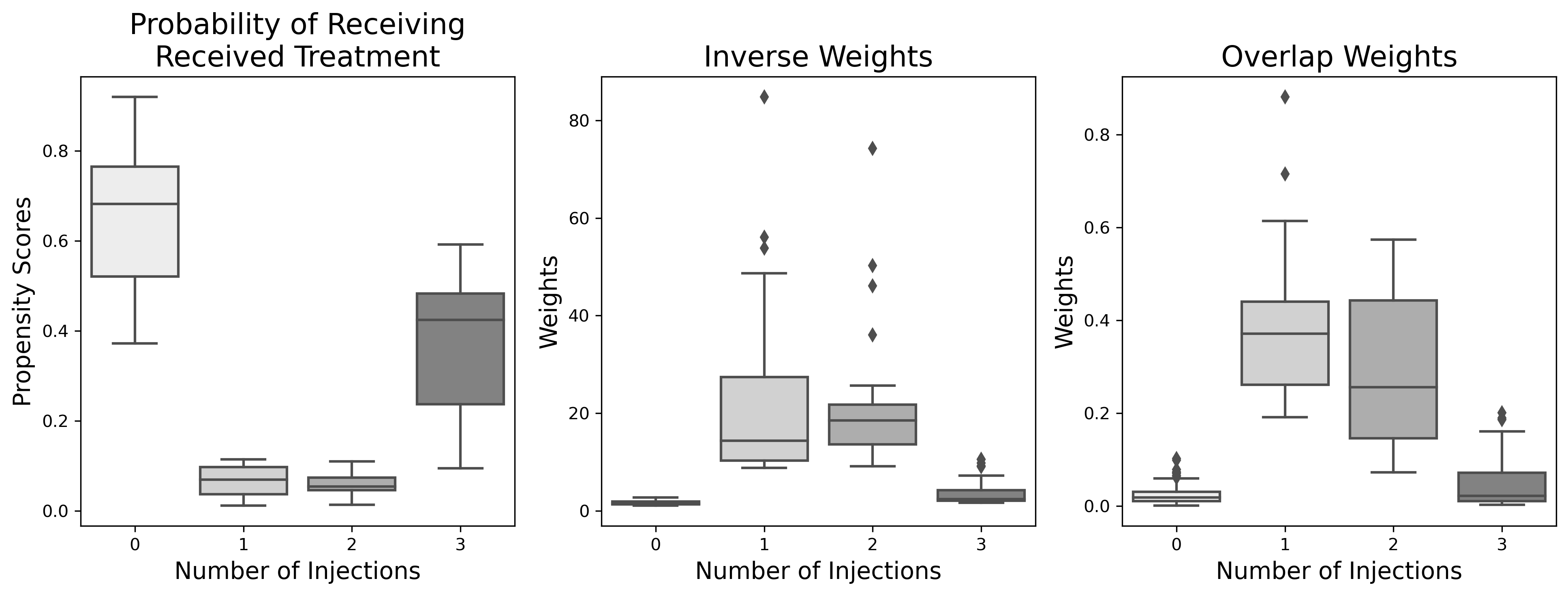}
  \caption{Boxplot of generalized propensity scores, inverse weights and overlap weights}
  \label{fig:propensity-weights}
\end{figure*}

As expected, participants have a higher likelihood to receive 3 or no injections compared to receiving 1 or 2 injections in an injection cycle. This follows immediately from the small sample size of patient-stages where only 1 or 2 injections were administered as a cycle. Likewise, in the boxplots of IPT and overlap weights, it is shown that observations associated with treatment groups $A =$ 1 or $A =$ 2 receive larger weight values in the dWOLS analysis to accommodate for their underrepresentation in the dataset.

\pagebreak
\section*{Appendix G: Residual Plots for Outcome Variable \texorpdfstring{$U(0.7)$}{LG}}

\begin{center}
    \begin{figure*}[!htp]
      \begin{minipage}{.45\textwidth}
        \centering
        \includegraphics[width=\linewidth]{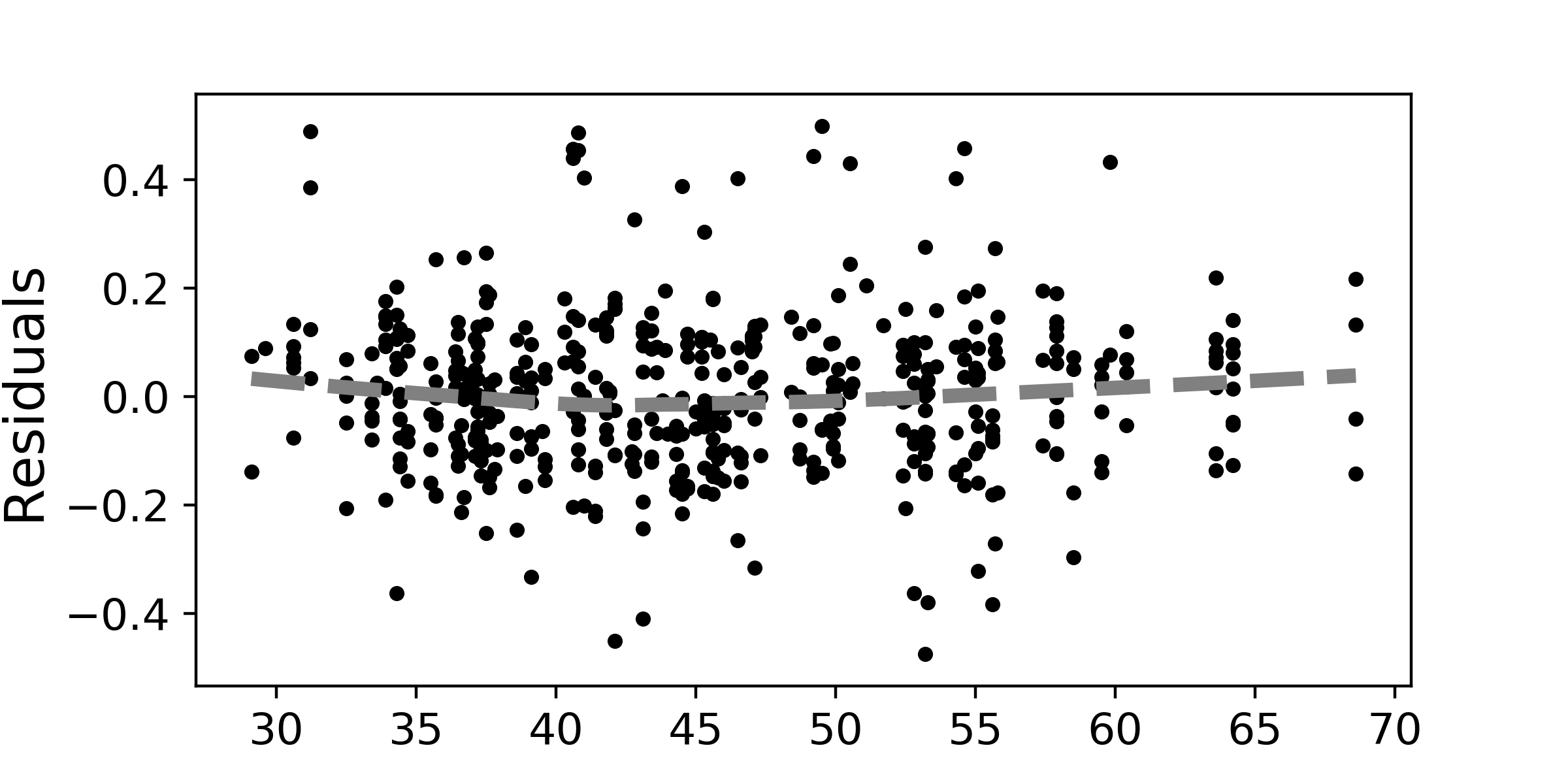}
      \end{minipage}
      \begin{minipage}{.45\textwidth}
        \centering
        \includegraphics[width=\linewidth]{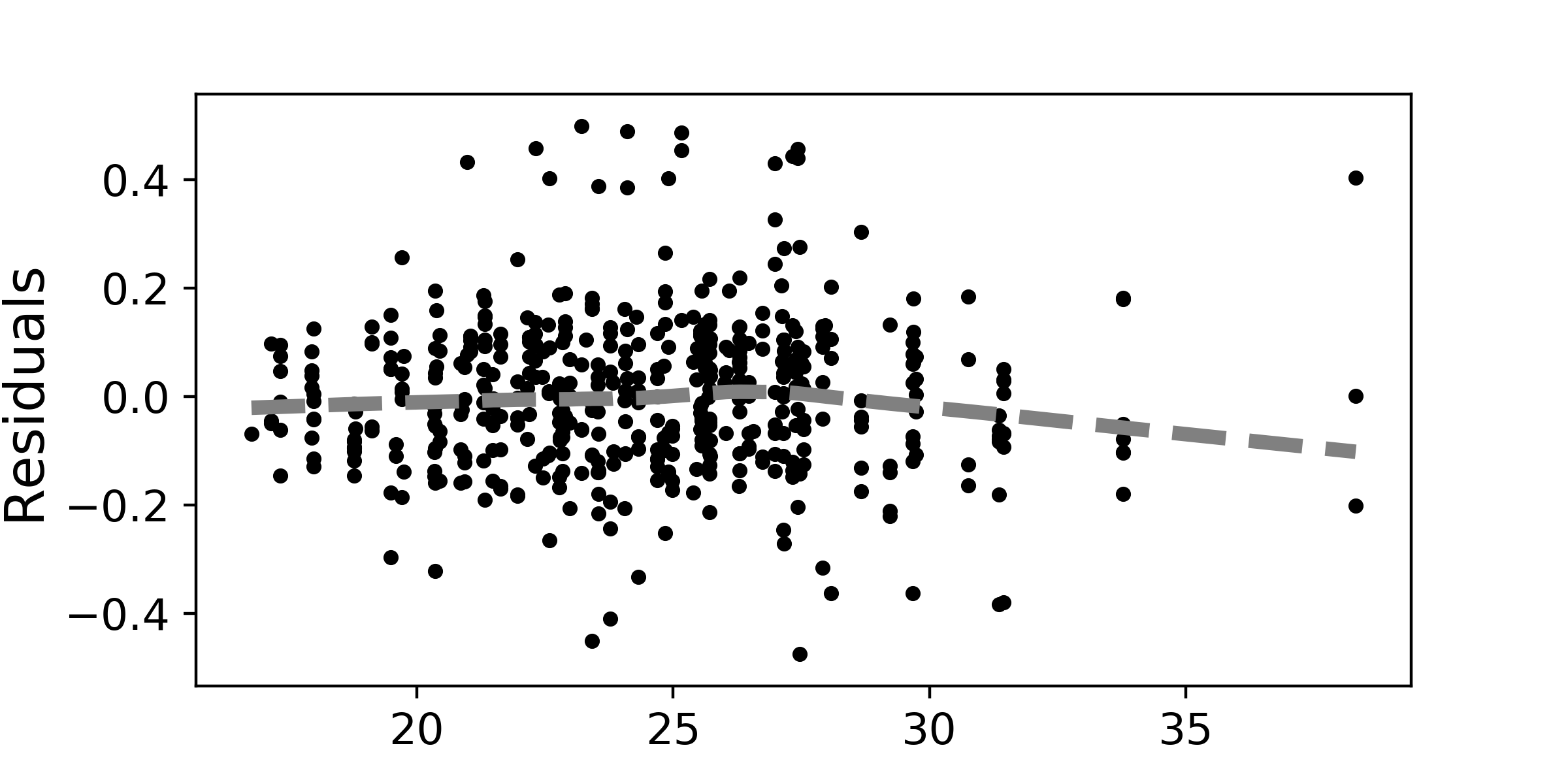}
      \end{minipage}\\
      \begin{minipage}{.45\textwidth}
        \centering
        \includegraphics[width=\linewidth]{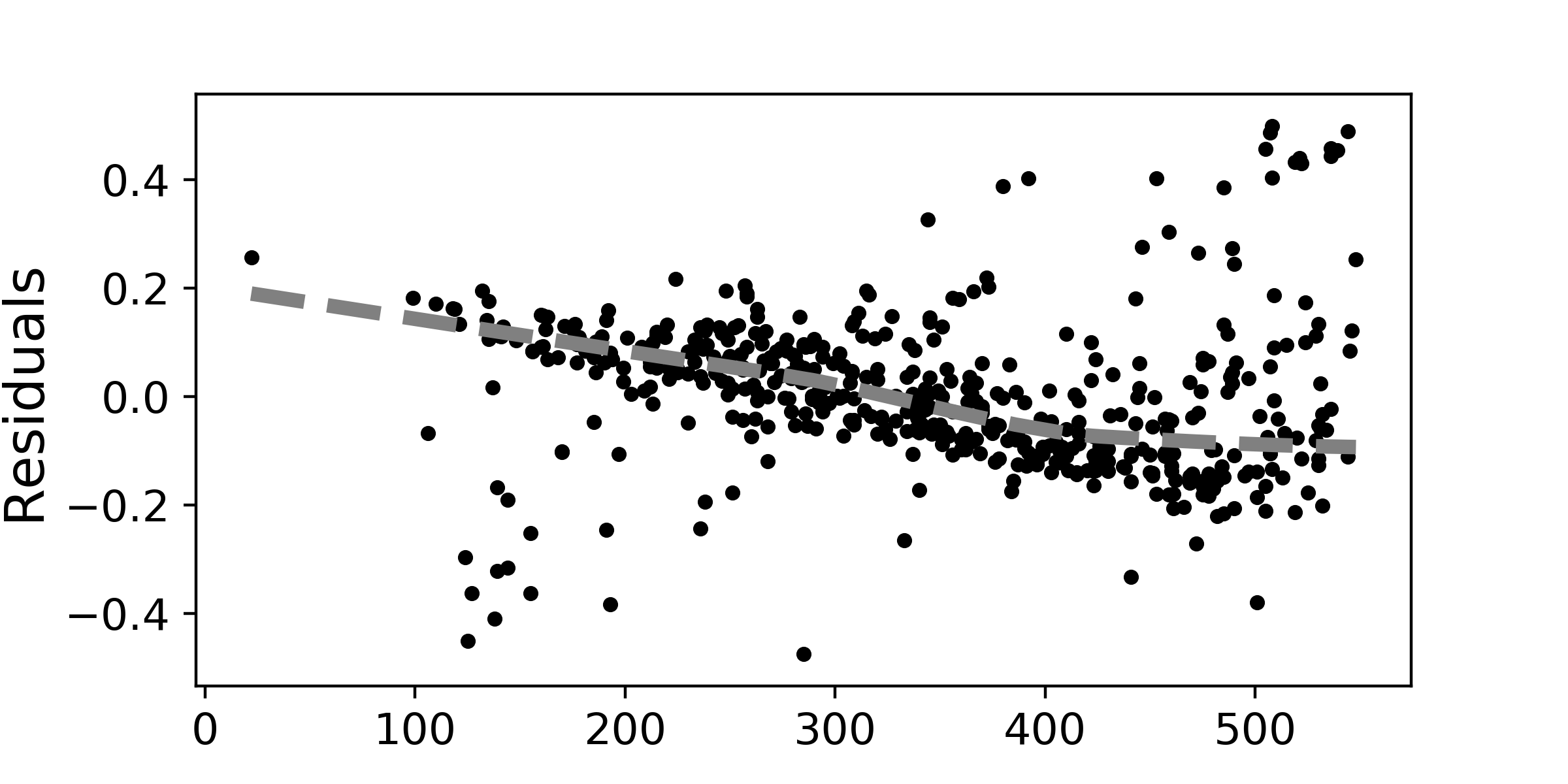}
      \end{minipage}
      \begin{minipage}{.45\textwidth}
        \centering
        \includegraphics[width=\linewidth]{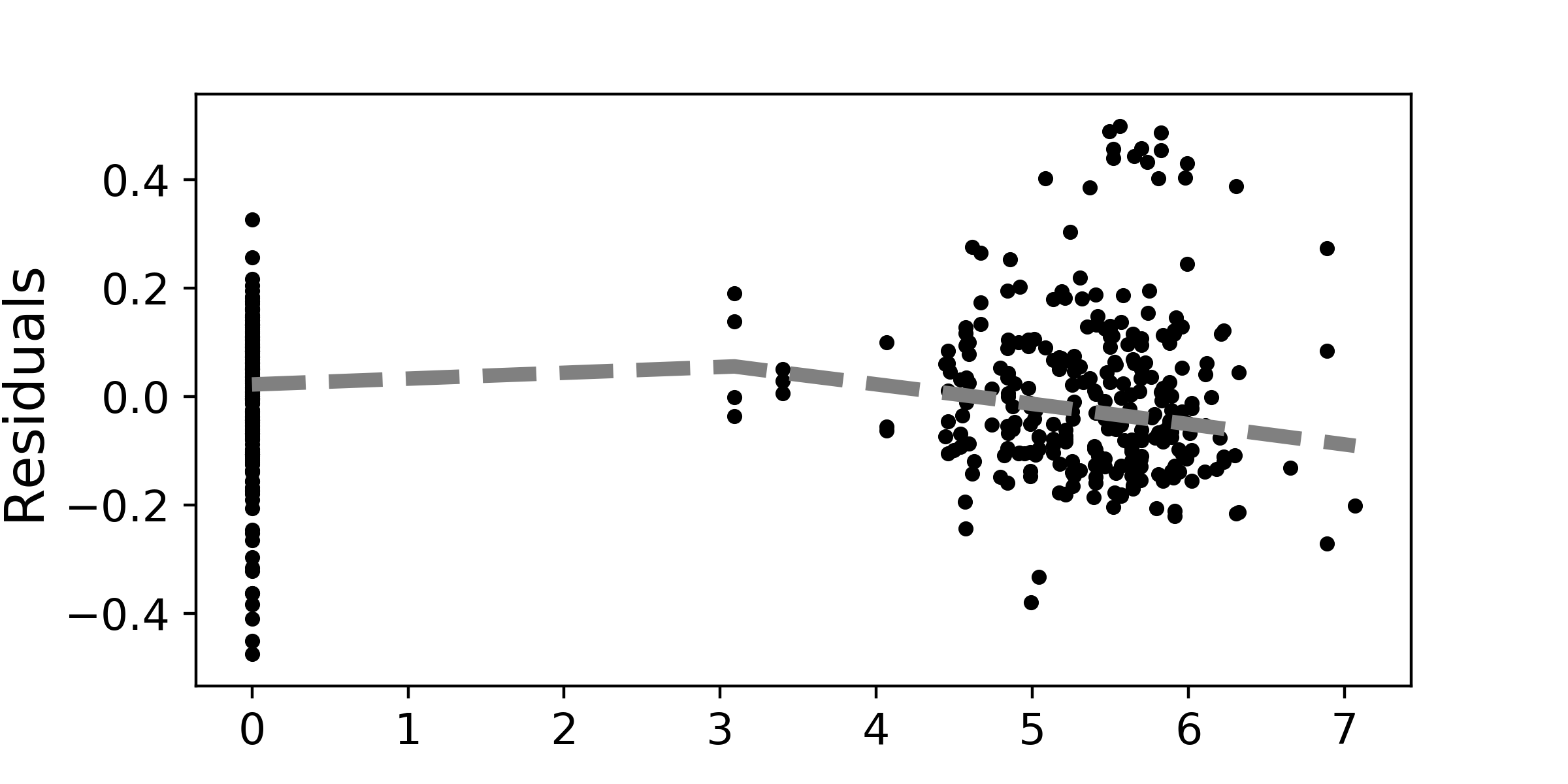}
      \end{minipage}\\
      \begin{minipage}{.45\textwidth}
        \centering
        \includegraphics[width=\linewidth]{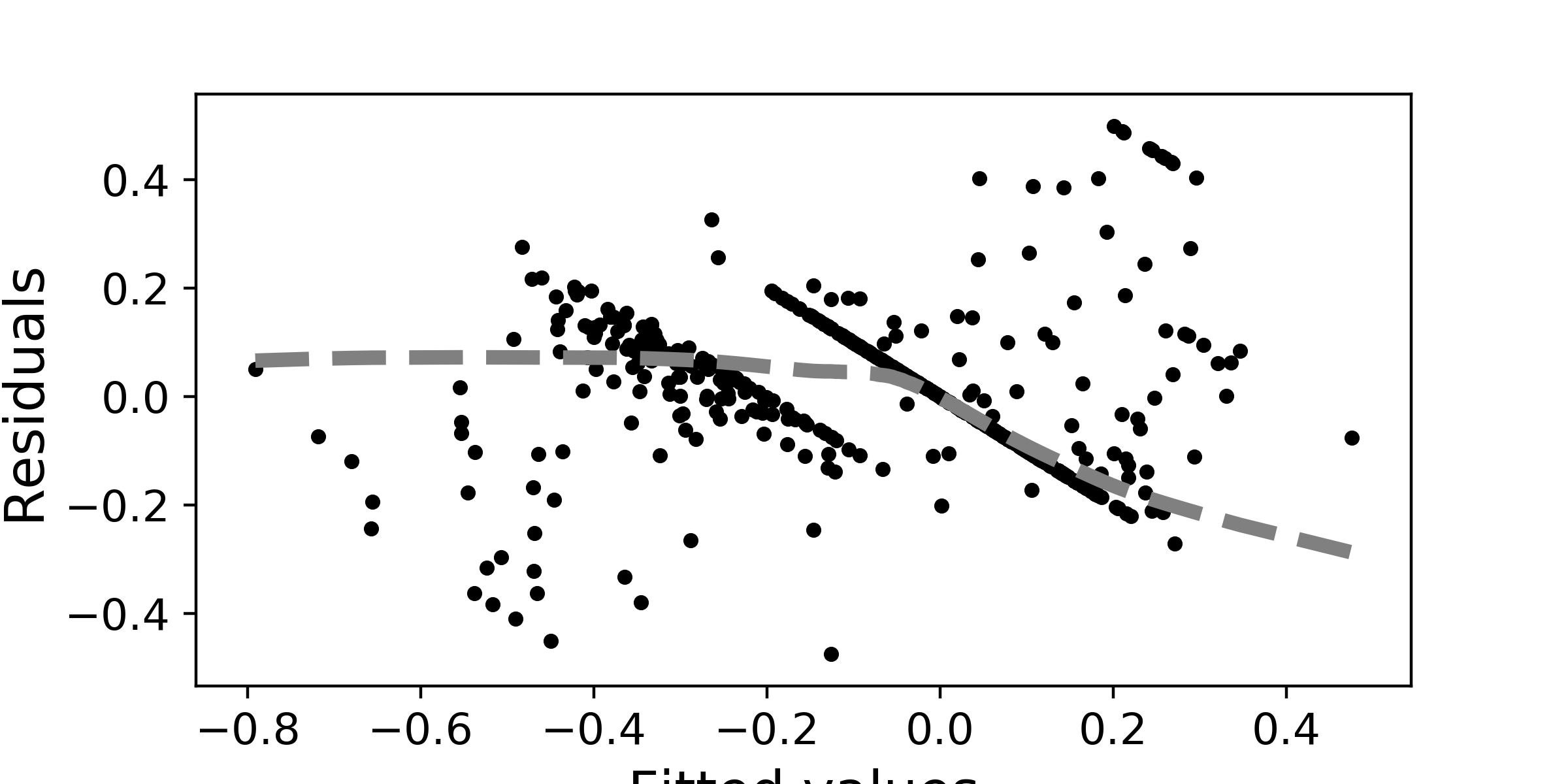}
      \end{minipage}
      \caption{Residuals plotted versus Age, BMI, $\text{CD4}_{\init}$, logResp and fitted values for outcome variable $U(0.7)$}
    \end{figure*}
\end{center}

\pagebreak
\section*{Appendix H: Residual Plots for Outcome Variable \texorpdfstring{$U(0.9)$}{LG}}

\begin{center}
    \begin{figure*}[!htp]
      \centering
      \begin{minipage}{.45\textwidth}
        \centering
        \includegraphics[width=\linewidth]{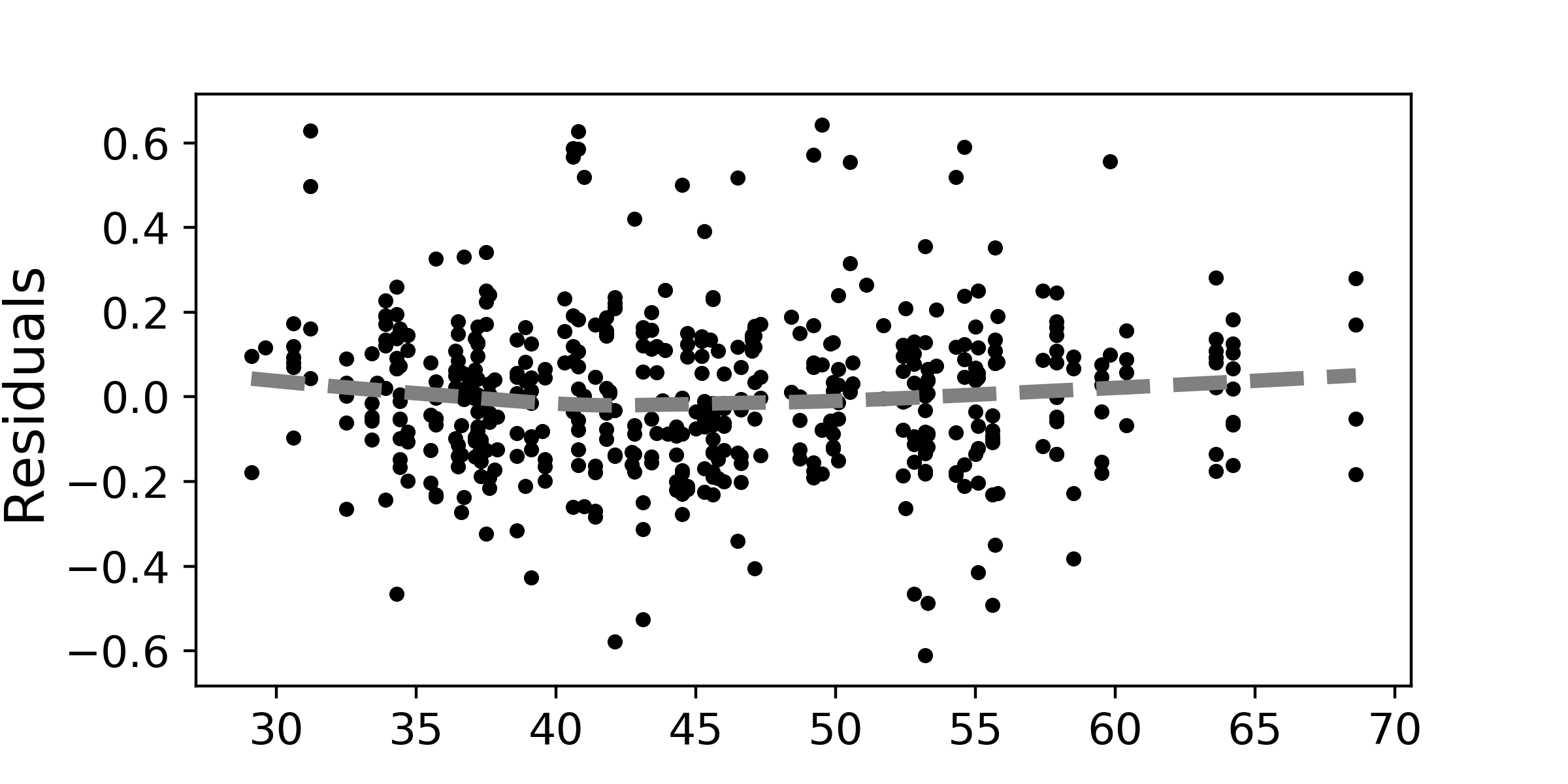}
      \end{minipage}
      \begin{minipage}{.45\textwidth}
        \centering
        \includegraphics[width=\linewidth]{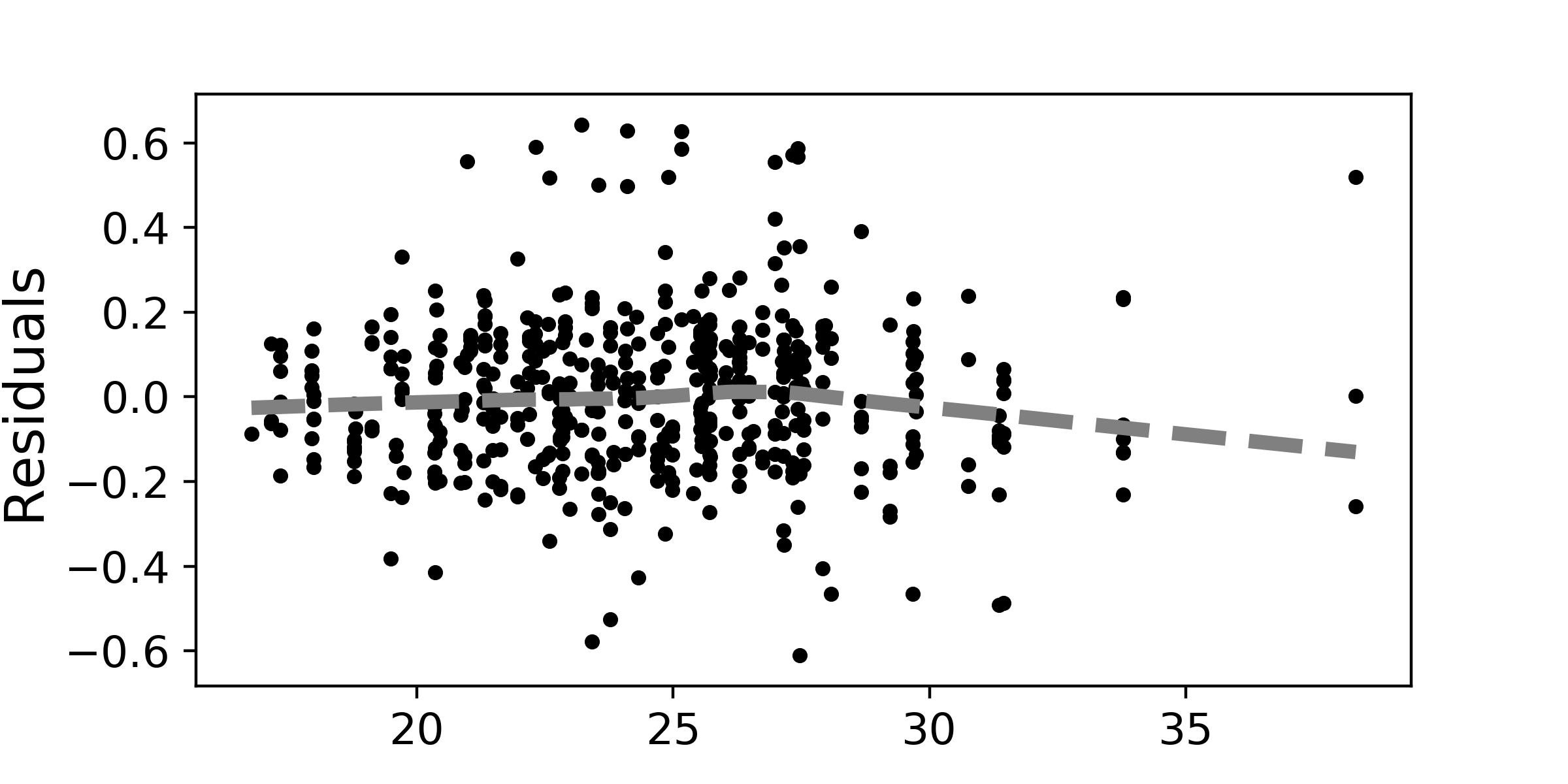}
      \end{minipage}\\
      \begin{minipage}{.45\textwidth}
        \centering
        \includegraphics[width=\linewidth]{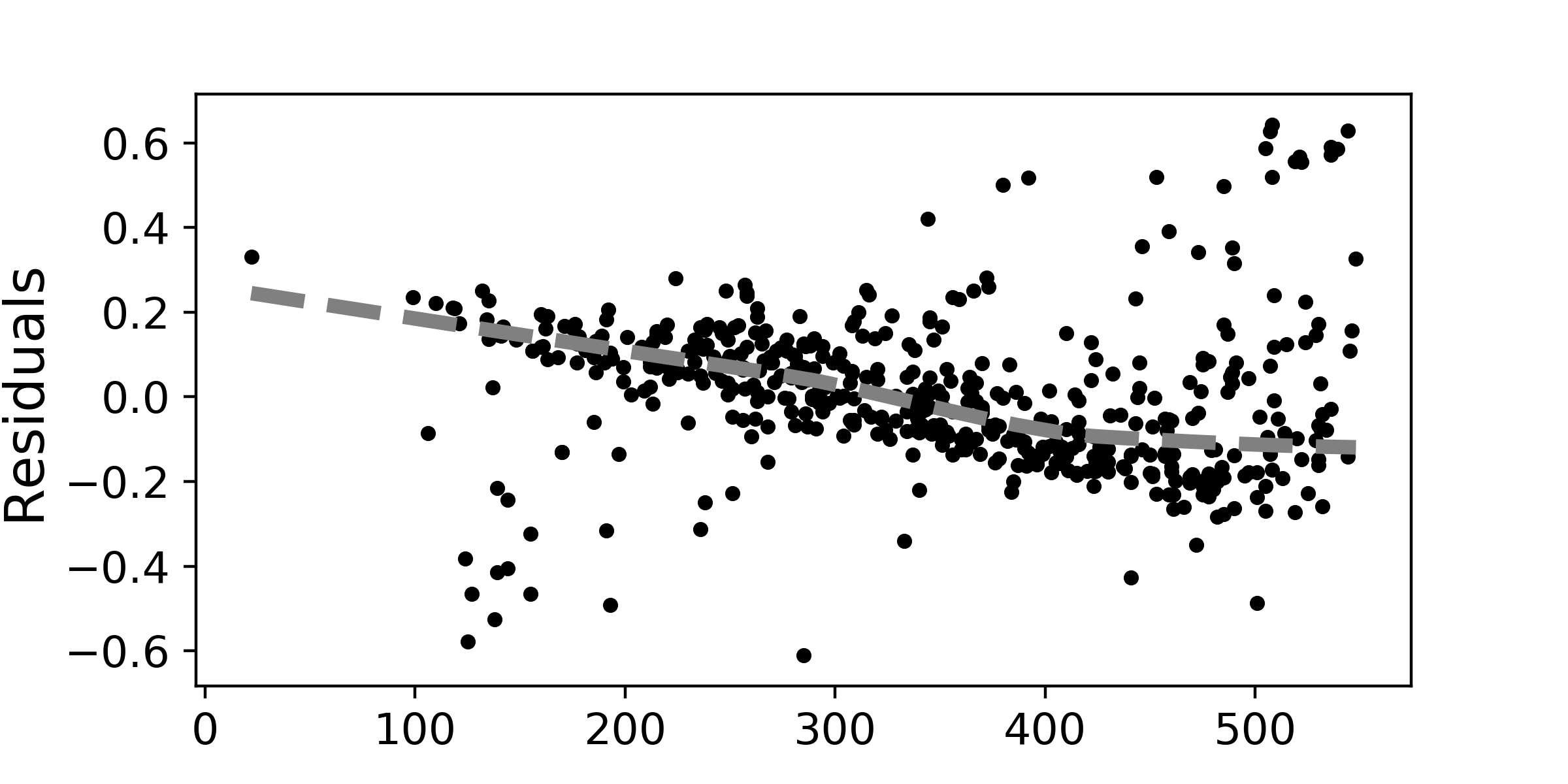}
      \end{minipage} 
      \begin{minipage}{.45\textwidth}
        \centering
        \includegraphics[width=\linewidth]{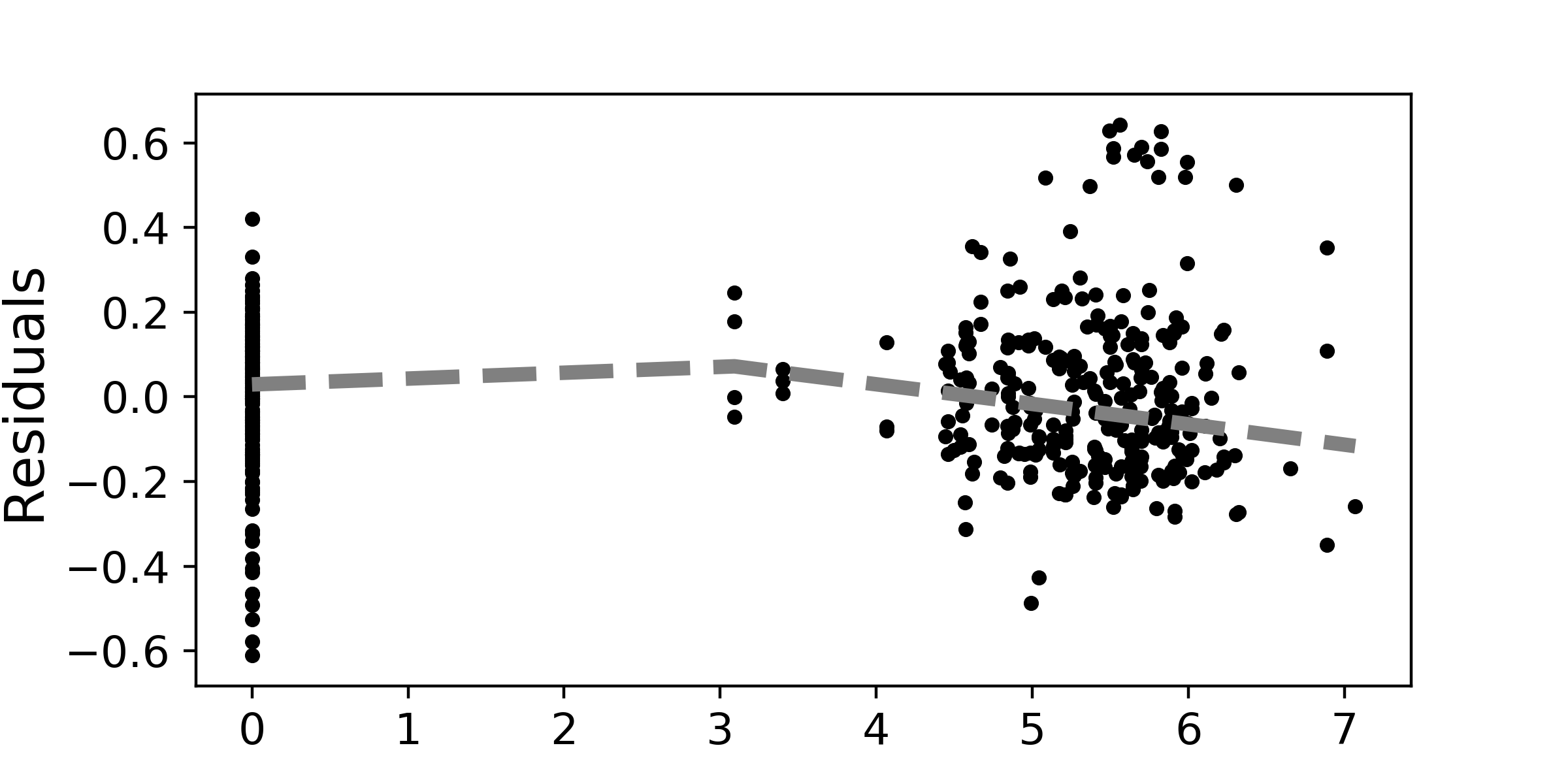}
      \end{minipage}\\
      \begin{minipage}{.45\textwidth}
        \centering
        \includegraphics[width=\linewidth]{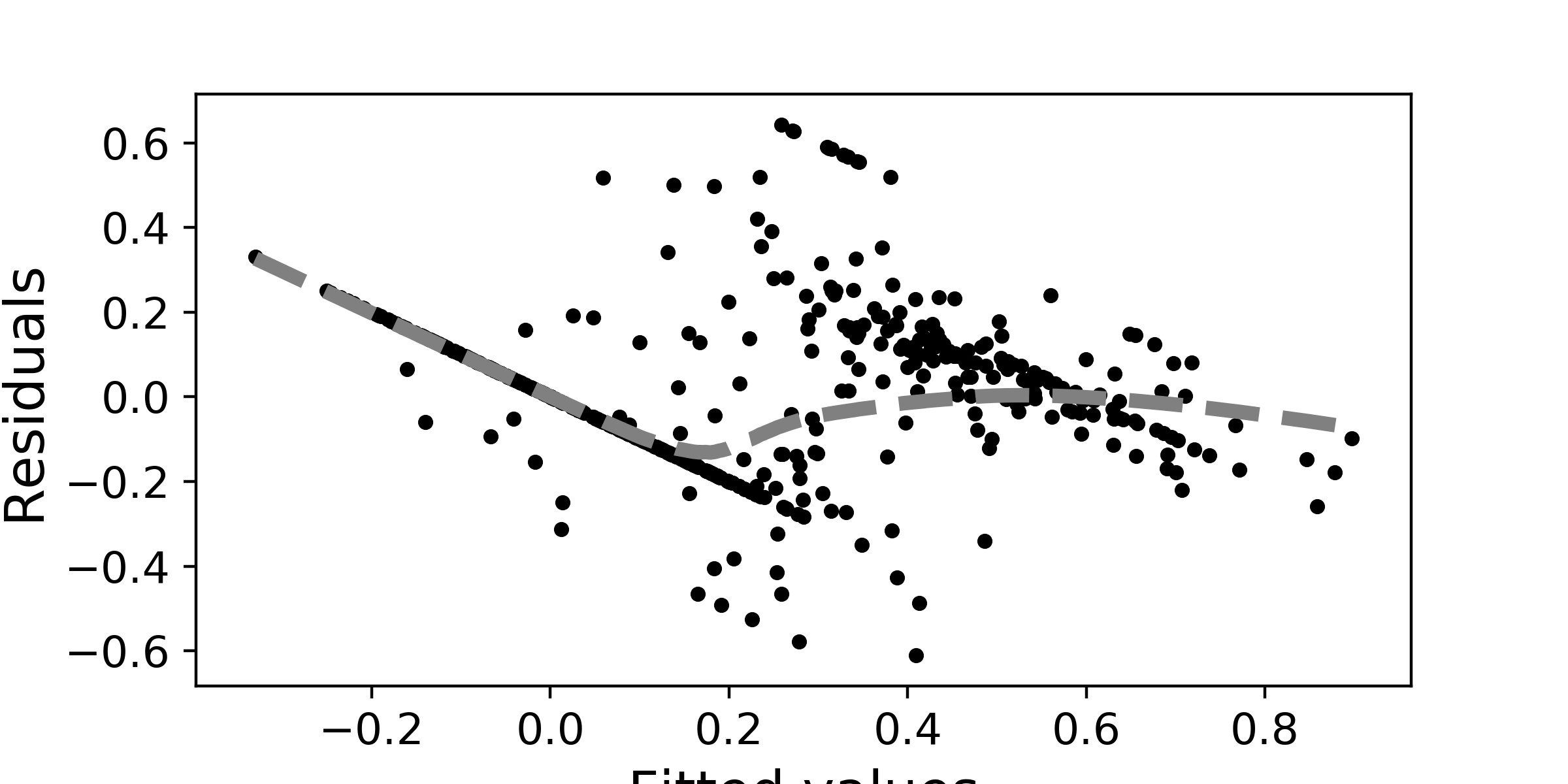}
      \end{minipage}
      \caption{Residuals plotted versus Age, BMI, $\text{CD4}_{1}$, logResp and fitted values for outcome variable $U(0.9)$}
    \end{figure*}
\end{center}

\end{document}